\documentclass[showpacs,preprintnumbers,amsmath,amssymb,prd]{revtex4}
\usepackage{graphicx}
\usepackage{epsfig}
\usepackage{dcolumn}
\usepackage{bm}

\begin{document}

\title{The Bispectrum and the trispectrum of the Ostriker-Vishniac effect}

\author{P. G.~Castro\footnote[2]{pcastro@astro.ox.ac.uk;pgc@roe.ac.uk}}
\affiliation{Astrophysics, Denys Wilkinson building, University of Oxford, Keble Road, 
Oxford OX1 3RH, U.K. \\ Phys. Rev. D 67, 123001 (2003)}

\begin{abstract}
We present analytical expressions for the Fourier analog of the CMB 
three-point and four-point correlation functions, the spatial bispectrum 
and trispectrum, of the Ostriker and Vishniac effect in the linear 
and mildly nonlinear regime. Through 
this systematic study, we illustrate a technique to tackle 
the calculation of such statistics making use of the effects 
of its small-angle and
vector-like properties through the Limber approximation. Finally we 
discuss its configuration dependence and detectability in the context 
of Gaussian theories for the currently favored flat $\Lambda$CDM cosmology.
\end{abstract}

\pacs{98.70.Vc}
\maketitle

\section{Introduction}
\label{sec:Introduction}
In recent years, with the prospect of the increase in the sensitivity 
and angular resolution of the forthcoming Cosmic Microwave Background 
(CMB) satellite and interferometry 
experiments~\cite{MAP/PLANCK,MINT,CBI,ACT}, efforts have been 
driven to the theoretical study of the secondary anisotropies contributions
 to the temperature fluctuations on arcminute scales and below. While 
the primordial anisotropies from recombination are thought to be well 
understood secondary anisotropies from reionization are not.

As is well known, the current favored inflationary model of structure 
formation predicts a nearly Gaussian probability distribution for the primordial 
density fluctuations. In this case, the CMB is completely described by its 
two-point correlation function or power spectrum in Fourier space. All 
higher-order correlations can be expressed in terms of it. Primordial
nonlinearities and secondary effects introduce deviations 
from Gaussianity, producing 
a detectable signal in both the power spectrum and higher-order statistics.
Recent work provides the theoretical background for the calculation
of estimators of these higher-order statistics~\cite{Hu01,Komatsu:Spergel00} 
and constrains possible non-Gaussian 
primordial contributions to the bispectrum and the trispectrum on degree 
and sub-degree angular scales using actual data 
~\cite{Kunz01,Santos01,Komatsu01,Troia03}. The interest is now in 
forecasting the expected signals on smaller scales due to secondary 
anisotropies, checking whether they are detectable and understanding how they 
can be separated from each other and from the primary anisotropies in light 
of the future data.

The Ostriker and Vishniac (OV) effect~\cite{Ost:Vish86} was 
found to be the dominant linear secondary contribution to the CMB 
anisotropies below the Silk-damping scales at the arcminute level~\cite{Hu94}. It is
caused by Thomson scattering off of the CMB photons by moving electrons 
during the initial phase of reionization. It has the advantage of taking place 
during the linear regime of structure evolution and of being a 
small-angle effect enabling one to obtain analytical expressions for its 
higher-order correlation functions in the small-angle limit. Because of the highly 
predictive power of linear theory, any measurement of such statistics would 
be a sensitive probe of the reionization history of the universe, 
difficult to disentangle in measurements from nonlinear contributions. 
Several derivations for its power spectrum 
have been carried out~\cite{Hu94,Dod:Jubas93,Vishniac87} 
but no one has fully addressed the calculation of its bispectrum 
or trispectrum.

It is then timely to obtain these expressions and to qualify and quantify them. 
We therefore extend and detail previous techniques used for 
the calculation of the OV power spectrum to the calculation of its higher-order 
statistics. As it will be shown, for the particular case of fields which are 
vector-like in nature, such as the OV effect, even moments will dominate over 
odd moments, making the trispectrum a much more sensitive statistics than 
the bispectrum. 

Given the low redshifts of formation of structure, it is interesting to 
consider whether nonlinear effects can further enhance these statistics. So we will 
extend our study to the weakly nonlinear regime, allowing us to 
probe the most natural extension of the OV effect to nonlinear scales, 
the so-called kinetic Sunyaev-Zel'dovich (kSZ) effect. On small scales, 
both arise from the density modulation of the Doppler effect from 
large-scale bulk flows.  

We review the relevant properties and parameters of the adiabatic 
cold dark matter (CDM) cosmology 
for structure formation in Sec.~\ref{subsec:Cosm}. In Sec.~\ref{subsec:OV} we review 
the theory of the OV effect and in Sec.~\ref{subsec:Stat} we discuss the basic 
statistical properties of a general field through its n-point functions. 
In Sec.~\ref{subsec:DomCont} we show how the homogeneous theory of turbulence 
combined with the Limber approximation enables one to infer the dominant 
contribution among n-point statistics of a vector field effect like the 
OV effect. In Sec.~\ref{subsec:Signal-to-noise}, we introduce the standard
formalism of the signal-to-noise calculation by means of the Fisher matrix. 
For general consistency, in Sec.~\ref{subsec:PowSpectrum} we 
apply our method in detail to the calculation 
of the OV power spectrum as well as its nonlinear extension.
In Sec.~\ref{sec:Bispectrum} 
and in Sec.~\ref{sec:Trispectrum} we present the steps of the calculation of the 
OV bispectrum and of the trispectrum respectively and its nonlinear counterparts.
We also present the results. Finally in Sec.~\ref{sec:Conclusions} we conclude.
In the Appendix A we generalize the Limber approximation to the 
3-point and 4-point correlation functions. This may be useful 
for other cosmological studies.

\section{General Considerations}
\label{sec:GenCons}
\subsection{Cosmological model} \label{subsec:Cosm}
We work in the context of the adiabatic cold dark matter (CDM)
family of models. In units of the critical density, 
$\Omega_0$ is the contribution to the non-relativistic-matter
density, $\Omega_b$ is the contribution to the baryonic matter
density, $\Omega_\Lambda$ is the contribution of the 
cosmological constant and $H_0 = 100\, h$ km~sec$^{-1}$~Mpc$^{-1}$ 
is the Hubble constant today.
The Friedmann equations for the evolution of the scale factor
of the Universe, $a(t)$, are then
\begin{equation}
     \frac{\dot a}{a} = H_0 E(z) \equiv H_0 \sqrt{\Omega_0 (1+z)^3 +
     \Omega_\Lambda + (1-\Omega_0-\Omega_\Lambda)(1+z)^2},
\end{equation}
\begin{equation}
     \frac{\ddot a}{a} = H_0^2 [\Omega_\Lambda - \Omega_0
     (1+z)^3/2],
\end{equation}
where the over-dot denotes a derivative with respect to
time. The scale factor is chosen such that $a_0 H_0=2c$.  

Useful measures of distance (and time) are the conformal distance (and
conformal time). If an observer is at the origin $z=0$ then an 
object at redshift $z$ is at a comoving distance, 
$w(z) = {1\over 2} \int_0^z \, {dz' \over E(z')}$
and at a time
$t(z) = {1 \over H_0} \int_z^\infty \, {dz' \over (1+z') E(z')}$.
The conformal time is then obtained from $d\eta=dt/a$, such that the comoving
distance to the horizon is the conformal time today, $c\eta_0=w(\infty)$.

If the CDM density contrast at comoving position $\vec{w}$ at time $t$
is $\delta(\vec{w},t)$, then the power spectrum $P(k,t)$ is defined by 
the expectation value over all realizations
$<\tilde \delta(\vec{k},t) \tilde \delta^*(\vec{k}',t)> = (2\pi)^3
 \delta_D^3(\vec{k}-\vec{k}') P(k,t)$
where $\delta_D^3$ is the Dirac delta function.
In linear theory, $\delta(\vec{w},t)=\delta_0(\vec{w})D(t)/D(t_0)$,
where $t_0$ is the age of the Universe,
$\delta_0(\vec{w}) \equiv \delta(\vec{w},t_0)$, and the growth factor,
as a function of redshift, is~\cite{Peebles}
\begin{equation}
     D(z) = {5 \Omega_0\, E(z) \over 2} \, \int_z^\infty \, {1+z' \over
     [E(z')]^3} \, dz'.
\end{equation}
The power spectrum is given by $P(k,t)=P(k)(D/D_0)^2$, where
$D_0\equiv D(t_0)$. For $P(k) \equiv P(k,t_0)$ we use
\begin{eqnarray}
     P(k) &=& {2 \pi^2 \over 8}\, \delta_H^2\, (k/2)^n\, T^2(k_p\,{\rm
     Mpc}/h \Gamma),
\label{eq:powerspectrum}
\end{eqnarray}
where $T(q)$ is the CDM transfer function, $k_p= k a_0 =k
H_0/2c$ is the physical wave number with our conventions and $\Gamma$,
the shape parameter, is defined as~\cite{Hu:Sugiyama96}
$\Gamma \simeq \Omega_0(h/0.5)\exp(-\Omega_b-\Omega_b/\Omega_0)$.
As we chose $a_0H_0=2c$, there is an extra factor of 8 in the denominator in
Eq.~(\ref{eq:powerspectrum}). For the transfer function, we use the
Bardeen \emph{et al.}~\cite{Bardeen86} fitting formulae for CDM models
instead of the improved version of Eisenstein and Hu~\cite{Eisenstein:Hu99}, 
to facilitate comparison with previous work.
For $\delta_H$, we take the fits to the Cosmic Background Explorer 
given in~\cite{Bunn:White97}. 

In linear theory, the continuity equation relates the Fourier 
components of the velocity field and the density field 
\begin{equation}
     \tilde{\vec{v}}(\vec{k},t) = {i a(t) \over k^2}\, {\dot D \over D}\, \vec{k}
     \, \tilde{\delta}(\vec{k}, t) = {i a(t) \over k^2}\, {\dot D
     \over D_0} \,\vec{k} \, \tilde{\delta}_0(\vec{k}).
\label{eq:veckeqn}
\end{equation}
A useful relation~\cite{Peebles} is
\begin{equation}
     {\dot D \over D} = {\ddot a \over \dot a} - {\dot a
     \over a} + {5 \Omega_0 \over 2} {\dot a \over a} {(1+z)^2
     \over [E(z)]^2\,D(z)}.
\end{equation}
Although we maintain generality in the derivations, we illustrate our results 
with a flat model, the $\Lambda CDM$ model. The parameters for this model 
are $\Omega_0=0.35$, $\Omega_b=0.05$, $\Omega_{\Lambda}=0.65$, $h=0.65$ and 
spectral index $n=1$. 
Concerning the reionization contribution we consider two 
reionization histories, both assuming steep reionization with 
ionization fraction $x_e=1$ and $\Delta z_r/(1+z_r)=0.1$. 
In the first one, reionization takes place at $z_r=8$. 
The second one assumes $z_r=17$
and relies on the latest results from the Wilkinson Microwave Anisotropy Probe 
(WMAP) experiment (see below).
Note that in an open or closed universe one replaces 
in the factors of $\eta$ that appear in the equations
\begin{equation}
  \eta \rightarrow \frac{ S(a_0 H_0 \eta \sqrt{ \mid 1 - \Omega_0 - \Omega_{\Lambda} \mid }) }
                     { a_0 H_0 \sqrt{ \mid 1 - \Omega_0 - \Omega_{\Lambda} \mid }}
\label{eq:transferfunction}
\end{equation}
where $S(x)=\sinh{x}$ in an open universe and $S(x)=\sin{x}$ in a closed 
universe.
\subsection{The Ostriker-Vishniac effect} \label{subsec:OV}
The reionization of the Universe is one of the most important physical 
processes that took place in the early universe
(see~\cite{Barkana:Loeb01,Haiman:Knox99}). The most accepted sources 
for reionization, which requires a source of ultra-violet photons, are an 
early generation of massive stars formed in dwarf galaxies or an early 
generation of quasars/AGNs in galaxies.
In the currently favored adiabatic CDM 
class of models for structure formation, reionization is expected to occur 
in the range $8 \leq z_r \leq 30$. Measurements of the CMB anisotropies
on sub-degree scales~\cite{Bernardis00,Hanany00,Netterfield01} 
have been used to put an upper bound on the 
reionization redsfhit of $z_r \simeq 30$~\cite{Griffiths99,Tegmark00}.
Very recently, using polarization and temperature anisotropies of the CMB,
WMAP has placed a fairly model-independent 
constrain~\cite{Hu:Holder03} on the optical depth to 
electron scattering of $\tau = 0.17 \pm 0.04$ at $68$\% CL 
which translates into $z_r=17 \pm 3$ for instant 
reionization~\cite{Kogutetal03,Spergeletal03}.
Interestingly, the measurement of an increase of the neutral fractions
with redshift in high-$z$ quasar spectra~\cite{Songaila:Cowie02,Djorgovski01}
and a first detection of the 'Gunn-Peterson trough'~\cite{Gunn:Peterson65} 
in a quasar spectra at $z = 6.28$ by SLOAN~\cite{Becker01} point to a 
reionization redshift of $z_r \simeq 6$~\cite{Gnedin01}, 
in disagreement with WMAP results. However, 
even a fraction of neutral hydrogen as small as $0.1$\% in the IGM could explain 
the result due to the large cross-section to Ly-$\alpha$ photons.
Together with the results from WMAP, high-$z$ quasar measurements
indicate that the reionization history is more complex than 
previously thought and attempts are being made to fully understand 
it~\cite{Haiman&Holder03,Hui&Haiman03,Wyithe&Loeb03}.
  
Though extensive analytical (for a complete derivation 
see~\cite{Hu94,Dod:Jubas93}) and numerical studies (see references below) 
have been done to try to quantitatively understand many of the effects 
originated by reionization on the CMB, accuracy is difficult to reach and 
uncertainties still remain. Reionization will 
leave multiple distinctive imprints on the CMB 
anisotropies by bringing the CMB photons and the free moving electrons into 
scattering contact again. Through that window, the low-$z$ period of
the universe evolution can be probed experimentally in more detail
from the appearance of the first sources 
of ionization to the formation of the observable present large-scale 
structure. Studies have been done on the calculation of the 
contributions to the power spectrum of the CMB due to ionization 
induced effects like the Doppler effect on large angular 
scales~\cite{Sunyaev78,Kaiser84}, the thermal SZ effect and its kinetic 
analog~\cite{Bruscoli99,Ma:Fry01,Refregier:Teyssier00,Scaramella93,daSilva99,Springel00,Sunyaev:Zel80,Valageas00}, 
the Inhomogeneous 
Reionization~\cite{Aghanim95,Benson00,Gnedin00,Gruzinov:Hu98,Knox98} 
and the OV effect on smaller scales~\cite{Vishniac87,Hu:White96,Jaffe:Kam98}. 
Enlightening comparative studies between different effects 
can also be found~\cite{Gnedin:Jaffe00}.

As ionization effects introduce non-Gaussianities in the anisotropies,
further studies were done on the calculation of their possible
contributions to the bispectrum. Many 
authors~\cite{Goldberg:Spergel99,Cooray:Hu00,Cooray01a,Cooray01b,Komatsu02} 
investigated contributions to mixed bispectra due 
to couplings between lensing effects, Integrated Sachs-Wolfe effect, thermal SZ 
and Doppler effects, such as the OV effect. The trispectrum of 
ionization secondaries~\cite{Cooray01b}
hasn't been explored very much. No one has addressed 
the calculation of the pure bispectrum and trispectrum of the 
OV effect until now.

In the linear regime, for the power spectrum, the dominant 
small-angular scale contribution from 
reionization was found to be the OV 
effect~\cite{Ost:Vish86,Vishniac87}. It arises 
from the second order modulation of the Doppler 
effect by density fluctuations which affect the probability of scattering. 
Because of its density weighting, it peaks at small angular scales, typically 
arcminute scales in $\Lambda$CDM models, and should produce $\mu K$ 
anisotropies. Its contributions to the temperature 
fluctuations along the line of sight 
can be written in the manner of Jaffe and Kamionkowski (JK)~\cite{Jaffe:Kam98}
\begin{equation}
\frac{\Delta T}{\bar{T}}(\vec{\theta}) = - \int_{0}^{\eta_0}d\eta\,
    g(\eta)\,\hat{\theta}.{\mathbf p}(\hat{\theta}\eta,\eta)
\label{eq:ov_effect}
\end{equation}
where ${\mathbf p}(\hat{\theta}\eta,\eta) \equiv {\mathbf v}(\hat{\theta}\eta,\eta)\delta(\hat{\theta}\eta,\eta)$ and $g$ is the visibility function given by
\begin{equation}
g(\eta) = \frac{a(\eta)\,\sigma_T\,\bar{n_e}(\eta)}{c} e^{-\tau(\eta)}=\frac{0.138\,\Omega_b\,h}{c}(1+z(\eta))^2\,x_e(\eta)\,e^{-\tau(\eta)}
\label{eq:vis}
\end{equation}
which gives the probability of scattering. The prefactor $0.138$ is
obtained assuming that all the baryons are in the form of protons (if we 
use the fact that the mass fraction of helium is $25$\% then 
one should multiply it by $7/8$). The 
visibility function is normalized such that 
$\int_{0}^{\eta_0}g(\eta)d\eta=1-e^{-\tau_r}$ where $\tau_r$ is the 
optical depth to the surface of last scattering at recombination. Note 
that $g$ is only dependent on time, and not on position for the OV effect. 
The optical depth is given by 
$\tau(\eta)=\int_{0}^{\eta}c\,d\eta'\,\sigma_T\, n_e(\eta')$.
Also $\delta(\hat{\theta}\eta,\eta)$ and 
${\mathbf v}(\hat{\theta}\eta,\eta)$ are the density contrast 
and bulk velocity along the line of sight, $\bar{n_e}(\eta)$ is the mean 
electron density given by 
$\bar{n_e}(\eta)=\Omega_{b}\,\rho_{c}\,x_e(\eta)\,(1+z)^3/m_p$, 
$\sigma_T$ is the Thomson cross-section, $x_e$ is the ionization fraction and 
$m_p$ the proton mass. 
We assume that the visibility function is approximated 
as a Gaussian in conformal time
\begin{equation}
g(\eta) = \frac{1-e^{-\tau_r}}{\sqrt{2\pi(\Delta \eta_r)^2}}
          e^{-\frac{1}{2}\frac{(\eta-\eta_r)^2}{(\Delta \eta_r)^2}} .
\label{eq:visibility_Gaussian}
\end{equation}

Following JK~\cite{Jaffe:Kam98}, we choose a coordinate system such that 
$\hat{\theta}$ represents a three-dimensional unit vector along the line of 
sight and $\vec{\theta}$ refers to a two-dimensional unit vector in the plane 
perpendicular to the line of sight. So we will have 
$\vec{\theta}=(\theta_1,\theta_2,0)$ and 
$\hat{\theta}=(\theta_1,\theta_2,\sqrt{1-\theta_1^2-\theta_2^2})\simeq (\theta_1,\theta_2,1)$ where this approximation arises from the small-scale nature of the effect. 
Bold letters represent three-dimensional vectors. 

The OV is a small-angle effect so we can work under the flat-sky approximation 
and expand the temperature perturbations in Fourier space
\begin{eqnarray}
\frac{\tilde{\Delta T}}{\bar{T}}(\vec{K})
          &=&-\int_{0}^{\eta_0}d\eta g(\eta)\int d^2\theta \int_{}
               \frac{d^3q}{(2\pi)^3}\,
               \hat{\theta}.\mathbf{\tilde{p}}({\mathbf q},\eta)\,
               e^{i(\vec{K}.\vec{\theta}-\eta{\mathbf q}.\hat{\theta})}
\label{eq:FT}
\end{eqnarray}
where ${\mathbf q} \equiv (q_x,q_y,q_z)$, $\vec{K} \equiv (K_x,K_y,0)$ and
\begin{eqnarray}
\mathbf{\tilde{p}}({\mathbf q},\eta)
                        &=&\frac{ia(\eta)\mbox{\.{D}}D}{2D_0^2}\int_{}\frac{d^3{\mathbf k}}{(2\pi)^3}
                           \tilde{\delta}_0({\mathbf k})\tilde{\delta}_0({\mathbf q}-{\mathbf k})
                           \left(\frac{{\mathbf q}-{\mathbf k}}{|{\mathbf q}-{\mathbf k}|^2}
                                 + \frac{{\mathbf k}}{|{\mathbf k}|^2}   \right)
\label{eq:p}
\end{eqnarray}
is the Fourier transform of ${\mathbf p}(\hat{\theta}\eta,\eta)$ 
(see JK~\cite{Jaffe:Kam98}). $D$ and $\mbox{\.{D}}$ depend on $\eta$. We made 
use of the continuity equation in Fourier space (\ref{eq:veckeqn}) and of 
the linear evolution of the density field.
\subsection{Statistical properties of a general field} \label{subsec:Stat}
The statistical properties of a field can be characterized by the n-point 
correlation functions in real space or by the n-point spectra in Fourier space. 
If the field is Gaussian in nature, like the primordial density fluctuations 
field in the current favored inflationary cosmology, the connected part of the n-point 
functions disappears for $n>2$. The non-zero (even-n)-point 
correlation functions can be expressed with the 2-point correlation function. 
As a result, a Gaussian distribution is completely described by the two-point 
correlation function, or power spectrum, and any non-Gaussian field will be 
detectable by measuring the connected part of its n-point correlation function. 

If we consider a general statistically homogeneous and isotropic 2-dimensional 
field $\vec{p}$ with zero mean, its power spectrum $P$, bispectrum $B$ and 
trispectrum $T$ are defined by the following equations in the appropriate 
Fourier convention:
\begin{eqnarray}
<\vec{p}(\vec{k}_1)\vec{p}(\vec{k}_2)>&=& (2\pi)^2P(k_1)\delta_D^2(\vec{k}_1+\vec{k}_2)\nonumber\\
<\vec{p}(\vec{k}_1)\vec{p}(\vec{k}_2)\vec{p}(\vec{k}_3)>_c&=& (2\pi)^2B(k_1,k_2,k_3)\delta_D^2(\vec{k}_1+\vec{k}_2+\vec{k}_3) \label{eq:P-B-T}\\
<\vec{p}(\vec{k}_1)\vec{p}(\vec{k}_2)\vec{p}(\vec{k}_3)\vec{p}(\vec{k}_4)>_c&=&(2\pi)^2T(k_1,k_2,k_3,k_4)\delta_D^2(\vec{k}_1+\vec{k}_2+\vec{k}_3+\vec{k}_4)\nonumber 
\end{eqnarray}
where the subscript $c$ stands for connected. The OV effect being 
a secondary effect will introduce non-Gaussianities 
in the original primordial Gaussian distributed temperature fluctuations. 
As a consequence, contributions to its bispectrum and trispectrum are expected.
\subsection{Dominant contributions among the statistics of the OV effect}
\label{subsec:DomCont}
Combining the homogeneous theory of turbulence with the Limber 
approximation enables one to infer the dominant 
contribution among n-point statistics of an isotropic and homogeneous 
vector-like field effect
whose statistical properties vary slowly in time. In particular,
we can apply this to the OV effect.
In short, the theory of homogeneous turbulence shows how 
to build invariant spectral tensors of arbitrary order, corresponding 
to expectation values of arbitrary products of statistically homogeneous vector fields. 
It is based on techniques proposed in the area of 
homogeneous turbulence in the
1940s by Robertson~\cite{Robertson40}.
Relying on this theory, all 
expectation values of an odd product of an isotropic $3$-dimensional 
vector field 
${\mathbf p}({\mathbf q})$ with ${\mathbf q} \equiv (q_x,q_y,q_z)$ 
must be proportional to at least one 
of the ${\mathbf q}$ vectors, contrary to the expectation values 
of even products.

Because of the Limber approximation, extended to 
higher-statistics in Appendix A, which states that the 
only contributions to the projected correlation function on the sky 
come from the Fourier modes perpendicular to the line of sight 
of the angular correlation 
function, all the $q_{iz}$ terms tend to be suppressed. There will be
different levels of suppression depending on the order 
of the $q_{iz}$ dependence of our statistics. 

Combining these two results, we can conclude that even correlation
functions of the OV effect dominate over odd correlation functions 
making the trispectrum a much more sensitive statistics than 
the bispectrum. Also, we expect the correlation functions to obey
the homogeneous and isotropy theory fully and thus to be the result of 
contributions of different orders in the $q_{iz}$ terms. We have developed 
a method which permits to calculate the dominant contribution,
under the Limber approximation.

This is a characteristic of all effects physically described by an
isotropic vector field and can thus be useful for other studies. 
As noted previously by Scannapieco~\cite{Scannapieco00}, the alternation of 
dominant/subdominant/dominant higher order correlation functions provides 
a unique signal distinguishing the OV effect from other 
non-vector like secondary anisotropies and consequently
can enable one to disentangle it from other
contributions at similar angular scales. 
\subsection{Signal-to-noise} \label{subsec:Signal-to-noise}
A fundamental issue is to know how well we can separate the 
OV signal, which is non-Gaussian, from the Gaussian 
signal, noise and foregrounds which are always present in
a measurement from an experiment.
A way of quantifying this detection is to calculate the 
$\chi^2$ statistics (as in~\cite{Goldberg:Spergel99,Spergel:Goldberg99}).
To do so, one needs to calculate the Fisher information
matrix $F_{ij}$ (for a good review see~\cite{Tegmark96}). 
If we think of the data ${\mathbf x}$ as a random
variable with a likelihood function $L({\mathbf x};\theta)$ where
$\theta$ is a vector of model parameters, the Fisher information matrix
is defined as
\begin{equation}
        F_{ij} \equiv - 
               \left<  
                     \frac{\partial^{2}ln L({\mathbf x};{\mathbf \theta})}
                          {\partial \theta_i \partial \theta_j} 
               \right> .
\label{eq:fisher}
\end{equation}
By a very powerful theorem, called the Cramer-Rao 
inequality, it was shown~\cite{Kenney:Keeping51,Kendall:Stuart69} 
that the variance of any unbiased estimator of a certain 
parameter in a model, can not be less than $(F^{-1})_{ii}$. 
As the signal calculated is expected to be rather small, we are
interested in estimating its overall detectability as
in~\cite{Goldberg:Spergel99,Cooray:Hu00,Spergel:Goldberg99}. Therefore, 
we assume that the form of our model ${\mathbf \theta}$ (in our case
the bispectrum and the trispectrum)
is correct and that the 
only interesting parameter is its amplitude $A$, where 
the true value of $A = 1$. Then the Cramer-Rao inequality tells 
us that the variance of the measurement of $A$ is no less
than $\sigma^2(A)= (F^{-1})_{AA}$ and we define the 
$\chi^2$ statistics as
\begin{equation}
\chi^2 \equiv  \left( \frac{S}{N} \right)^2
        = \frac{ 1 }{\sigma^2(A)}
        = (F)_{AA} .
\label{eq:chi}
\end{equation}
The calculation of the Fisher matrix $(F)_{AA}$ 
of the statistics of the OV effect
involves the calculation of the contribution of the noise to 
the power spectrum $C_{\ell}^{noise}$, as it will be shown. The noise 
depends on the experiment characteristics.\vspace{4mm}

We consider a hypothetical experiment which maps a fraction of 
the sky $f_{sky}$ with a Gaussian beam with full width at half
maximum $\theta_{fwhm}$ and pixel noise 
$\sigma_p=s / \sqrt{t_{pix}}$, where $s$ is the detector sensitivity
and $t_{pix}$ is the time spent observing each pixel. We use the
inverse weight per solid angle, 
$w^{-1} \equiv (\sigma_p\theta_{fwhm}/T_0)^{2}$, in order to have
a measure which is independent of the pixel-size~\cite{Jungman96,Knox95}. 
$T_0=2.73 K$ is the CMB thermodynamic temperature. 
If only a fraction $f_{sky}$ of the sky is mapped, treating 
the pixel noise as Gaussian and ignoring 
any correlations between pixels a good estimate 
of the $C_{\ell}^{noise}$~\cite{Cooray:Hu00,Jungman96,Knox95} is 
\begin{equation}
C_{\ell}^{noise}=f_{sky}\,w^{-1}\,e^{\sigma_b^2\,\ell\,(\ell+1)}
\label{eq:cl_noise}
\end{equation}
where $\sigma_b$, in radians, is the width of the beam 
if we assume it has a Gaussian profile. It is related to 
$\theta_{fwhm}$, in arcminutes, 
by $\sigma_b=\sqrt{8ln2}\,\theta_{fwhm}\times\pi/10800$.
Note that if an experiment maps the full sky and  
then a fraction $1-f_{sky}$ is subtracted, one should not
multiply $w^{-1}$ by $f_{sky}$ (case of MAP and Planck).
Hence we can estimate $C_{\ell}^{noise}$ for any experiment with 
characteristic $f_{sky}$, $\theta_{fwhm}$, $s$ and $t_{pix}$.

For the precise cases of MAP (renamed WMAP recently) and Planck, for which we used
the specifications in table~\ref{table1}, we need to take into account
their multi-frequency coverage with different characteristics. The
$C_{\ell}^{noise}$ is then defined as~\cite{Cooray:Hu00}
\begin{equation}
\frac{1}{C_{\ell}^{noise}}=\sum_{\nu} \frac{1}{C_{\ell}^{noise}(\nu)}
\label{eq:cl_noise_freq}
\end{equation}
where the sum runs over all channels of the experiment and $\nu$
is the frequency of the channel. 

\begin{table}[here]
\begin{tabular}[b]{c|@{\hspace{10mm}}ccc@{\hspace{10mm}}|@{\hspace{10mm}}ccc@{\hspace{10mm}}}
& \multicolumn{3}{c@{\hspace{10mm}}|@{\hspace{10mm}}}{MAP}& \multicolumn{3}{@{\hspace{-7mm}}c}{Planck} \\\hline
$\nu$ (GHz) &  $41$  &  $61$  & $95$ & 
               $100$ &  $143$ & $217$   \\
$\theta_{fwhm}$ (arcm)&  
	$31.8$ & $21.0$  & $13.8$ & 
	$10.7$ & $8.0$ & $5.5$  \\
$\sigma_p$ ($\mu$K)  & 
	$19.8$  & $30.0$ & $45.6$ &    
	$4.6$  & $5.4$  & $11.7$   \\\hline
$f_{\rm sky}$    & \multicolumn{3}{c@{\hspace{10mm}}|@{\hspace{10mm}}}{$0.80$} & \multicolumn{3}{@{\hspace{-7mm}}c}{$0.80$} \\
\end{tabular}
\caption{ \label{table1} Experimental parameters for (W)MAP and Planck.}
\end{table}
\subsection{Power spectrum of the OV effect} \label{subsec:PowSpectrum}
\subsubsection{Linear power spectrum}
In Fourier space, the flat-sky power spectrum of the OV effect is related 
(Eq.~(\ref{eq:P-B-T})) to the 
following two-point expectation value of the OV temperature field perturbation
$\Delta T/\bar{T}$
\begin{eqnarray}
	<\frac{\tilde{\Delta T}}{\bar{T}}(\vec{K_1})
 	\frac{\tilde{\Delta T}}{\bar{T}}(\vec{K_2})> 
               &=&     \frac{1}{2}\int_{0}^{\eta_0}d\eta_1 g(\eta_1) 
                       \int_{0}^{\eta_0}d\eta_2 g(\eta_2)
                       \int d^2\theta_1
                       \int d^2\theta_2 \nonumber\\
               & &     \int_{}\frac{d^3q_1}{(2\pi)^3}
                       \int_{}\frac{d^3q_2}{(2\pi)^3}\,
                       \hat{\theta}_{1i}
                       \hat{\theta}_{2j}\,
                      (<\tilde{p}_i({\mathbf q_1},\eta_1)
                        \tilde{p}_j({\mathbf q_2},\eta_2)>+
                       <\tilde{p}_i({\mathbf q_2},\eta_2)
                        \tilde{p}_j({\mathbf q_1},\eta_1)>)\nonumber\\
               & &     e^{ i(\vec{K_1}.\vec{\theta}_1-\eta_1{\mathbf q_{1}}\hat{\theta}_1) }
                       e^{ i(\vec{K_2}.\vec{\theta}_2-\eta_2{\mathbf q_{2}}.\hat{\theta}_2) }
\label{eq:pow_spect}
\end{eqnarray}
where $\tilde{{\mathbf p}}$ is defined as in Eq.~(\ref{eq:p}). 
Many authors have derived the expression for this statistics 
\cite{Hu94,Dod:Jubas93,Vishniac87,Jaffe:Kam98,Scannapieco00}. 
We use the JK formalism but a different 
technique which will be useful in what follows.

As we see, this expression involves a double integration in time, angle 
and wavenumber, being numerically long to evaluate. It is useful to note that as 
the statistical properties of the field ${\mathbf p}$ vary slowly in time 
and as the OV effect is a small-angle effect we can employ the Limber 
approximation (see A1) to considerably simplify our derivations. We stress
here that we are allowed to use this approximation as the OV effect takes 
place at sufficiently high $l$, where the difference between the 
approximation and the integral is very small.
 
As the two permutations $<\tilde{p}_i \tilde{p}_j >$ are symmetric due to statistical
homogeneity, we only consider the first one and multiply the result by $2$. 
Using Eq.~(\ref{eq:p}) for $\tilde{p}$ and the Wick theorem 
for the Gaussian 3-dimensional density field correlation function which states
\begin{eqnarray}
\lefteqn{ <\delta({\mathbf k}_1)\delta({\mathbf q}_1-{\mathbf k}_1)
           \delta({\mathbf k}_2)\delta({\mathbf q}_2-{\mathbf k}_2) > =}\nonumber\\
               & &    (2\pi)^6P(k_1)P(\mid {\mathbf q}_1-{\mathbf k}_1 \mid)
                      (\delta_D^3({\mathbf k}_1+{\mathbf q}_2-{\mathbf k}_2)
                       \delta_D^3({\mathbf q}_1+{\mathbf k}_2-{\mathbf k}_1)+
                       \delta_D^3({\mathbf k}_1+{\mathbf k}_2)
                       \delta_D^3({\mathbf q}_1-{\mathbf k}_1+{\mathbf q}_2-{\mathbf k}_2))
 \label{eq:Wick1}
\end{eqnarray}
where $P(k)$ is the power spectrum of density perturbations, we obtain two non-zero 
terms for $<\tilde{p}_i \tilde{p}_j>$ which can be written as
\begin{equation}
	<\tilde{p}_i({\mathbf q_1},\eta_1)
         \tilde{p}_j({\mathbf q_2},\eta_2)>=
          	       -2\,G(\eta_1)G(\eta_2)\,F_{ij}({\mathbf q}_1)\,
                       \delta^3_D({\mathbf q}_1+{\mathbf q}_2)
\label{eq:p_p}
\end{equation}
with the time dependence functions gathered in 
$G(\eta)=\left(\frac{iaD\dot{D}}{2D_0^2}\right)$ 
and the general tensorial functions $F_{\alpha\beta}$ given by
\begin{eqnarray}
F_{\alpha\beta}({\mathbf q}_i) &=& \int{}{} d^3K'P(a)P(b)
             \left( \frac{a_{\alpha}}{a^2}+\frac{b_{\alpha}}{b^2} \right)
             \left( \frac{a_{\beta}}{a^2}+\frac{b_{\beta}}{b^2} \right)
\label{eq:F_pow1}
\end{eqnarray}
where ${\mathbf a}={\mathbf K'}$ and ${\mathbf b}={\mathbf q}_i-{\mathbf K'}$.
We could now replace the two previous expressions directly in 
Eq.~(\ref{eq:pow_spect}) but, for clarity purposes as
it will become obvious soon, we refrain 
from doing so and instead we keep working with $F_{\alpha\beta}$.

Indeed, in the small-sky approximation, for which the unit vector 
$\hat{\theta} \simeq (0,0,1)$, we 
can contract the $\hat{\theta}$'s of expression (\ref{eq:pow_spect}) with 
the vectors ${\mathbf a}$ and ${\mathbf b}$ of the last expression (\ref{eq:F_pow1}) 
such that we are left with the line of sight components of ${\mathbf a}$ 
and ${\mathbf b}$. We can thus define a new scalar 
function $F$ such that 
\begin{eqnarray}
 F({{\mathbf q}}_i)=
           \hat{\theta}_{1\alpha}\hat{\theta}_{2\beta}F_{\alpha\beta} &\simeq& 
             \int{}{} d^3K'P(a)P(b)
             \left( \frac{a_z}{a^2}+\frac{b_z}{b^2} \right)^2
                   = \int{}{} d^3K'P(a)P(b)
             \left( \frac{a_z^2}{a^4}+\frac{2a_zb_z}{a^2b^2}+\frac{b_z^2}{b^4} \right).
\label{eq:F_pow2}
\end{eqnarray}
The interesting step that follows is to expand this function in $a_z=K'_z$ and 
$b_z=q_{iz}-K'_z$ 
\begin{eqnarray}
 F( {{\mathbf q}}_i ) = \int{}{} d^3K'P(a)P(b)
               \left(
                  K_z'^2\left( \frac{1}{a^4}+\frac{1}{b^4}-\frac{2}{a^2 b^2} \right)+ 
                  K_z'  \left( \frac{ 2q_{iz} }{a^2 b^2}-\frac{ 2q_{iz} }{b^4}  \right)+
                       \left( \frac{q_{iz}^2}{b^4} \right)
               \right).
\label{eq:F_pow3}
\end{eqnarray}
Expressions of this type will occur in the following and they 
illustrate the previous discussion in Sec.~\ref{subsec:DomCont}.
By looking at the integral, we see that the $K_z'$ integration 
for odd products of $K_z'$ is zero 
as the dependence on $K_z'$ of the terms with $a$ and $b$ is even. 
Therefore we are left with non-zero contributions 
from the first term (with $K_z'^2$) and 
from the last term (with $K_z'^0$). By applying 
the Limber approximation we can also infer that 
the dominant contribution has to 
come from the $K_z'^2$ term as it 
has no explicit dependency on $q_{iz}$. As we know, in 
the Limber approximation, Fourier modes parallel to 
the line of sight (terms on $q_{iz}$) tend to be suppressed.    
We are then left with $2$ terms which we will label dominant
and sub-dominant terms depending on the order of their 
cancellation. We are interested in the dominant one. This 
cancellation was forecasted in Sec.~\ref{subsec:DomCont}
in view of the homogeneous turbulence theory. Indeed,
the power spectrum of the OV effect is expected to have contributions only 
from terms with no dependency on $q_{iz}$ and with a $q_{iz}q_{jz}$ type dependency. 

After some straightforward algebra, in the Limber approximation framework, 
we find for the dominant contribution
\begin{eqnarray}
 F( {\mathbf q}_i ) &=& \int{}{} d^3K'P(a)P(b)
               \left(
                  K_z'^2\left( \frac{1}{a^4}+\frac{1}{b^4}-\frac{2}{a^2 b^2} \right) 
               \right)\nonumber\\
                            &=& -2\pi\, q_i 
                                 \int_{0}^{\infty}dy_1\int_{-1}^{1}d\mu
                                  P(q_i y_1)
                                  P(q_i y_2)
                                  \frac{(1-\mu^2)(1-2\mu y_1)^2}{y_2^4}
\label{eq:F_pow4}
\end{eqnarray}
where we have performed a spherical coordinate transform such that 
$\mu=\hat{q}_i.\hat{K'}$,
$a=y_1{q}_i$ and $b=y_2{q}_i$ with $y_2=\sqrt{1+y_1^2-2y_1\mu}$. 
To obtain the components of $K'_z$ in the
chosen coordinate system, we used the Limber approximation 
to assume $q_{iz} \simeq 0$, such that
$K'_z=K'\sqrt{1-\mu^2}$. This assumption preserves our dominant
term but suppresses any sub-dominant term
that could naturally arise when calculating the integral. 
As a consequence, in Eq.~(\ref{eq:F_pow3}), 
the dominant term (in $q_{iz}^0$)
may contain hidden contributions to the subdominant term (in $q_{iz}^2$). 
That this indeed is the case can be understood by a very 
simple reasoning. Consider Eq.~(\ref{eq:F_pow2}).
It is easy to show that the terms in $a_z^2$ and $b_z^2$ give 
identical contributions to the integral, such that if we 
calculate twice the integral in $a_z^2$ we should obtain 
the same result at the end, i.e., various terms depending on different
orders in $q_{iz}$. By doing this, our term in $q_{iz}^2$ present in 
Eq.~(\ref{eq:F_pow3}) simply disappears. We might then
expect it to show up in the integral calculation. But, 
most interestingly, when performing the calculations as previously using
the Limber approximation, we obtain the same result, i.e., 
Eq.~(\ref{eq:F_pow4}). As a consequence, by imposing $q_{iz}\simeq0$ in 
this example, we are in fact suppressing a sub-dominant term 
we know should be present. In conclusion,
there are more terms contributing to the sub-dominant
power spectrum than the one present in 
Eq.~(\ref{eq:F_pow3}) and these aren't so easily calculated. 
Hence we neglect all the sub-dominant terms in the 
Limber approximation. In the following, 
for the bispectrum and the trispectrum calculations,
similar problems are present but they will not affect the lowest order 
terms in $q_{iz}$ and even order in $K'_z$, which are the dominant terms
of interest to us.
\vspace{4mm}

We can finally combine Eqs.~(\ref{eq:pow_spect}),~(\ref{eq:p_p})
and~(\ref{eq:F_pow4}). Applying Kaiser's method and the Limber
approximation as described in the Appendix, we obtain the 
well-known expression for the dominant contribution
to the linear OV power spectrum
\begin{equation}
P^{OV}_{dom}(K)=\frac{1}{8\pi^2}\int_{0}^{\eta_{0}}\frac{g^2(\eta)}{\eta^2}[a(\eta)]^2
                    \left( \frac{ {\mbox{\.{D}}}D}{D_0}\right)^2S(K/\eta)d\eta
\label{eq:ov_effect_JK}
\end{equation}
where
\begin{equation}
S(k)=k\int_{0}^{\infty}dy_1\int_{-1}^{1}d\mu 
                                        P(ky_1)P(k\sqrt{1+y_1^2-2y_1\mu})
                                        \frac{(1-\mu^2)(1-2\mu y_1)^2}
                                              {(1+y_1^2-2\mu y_1)^2} .
\label{eq:S}
\end{equation}
The Limber approximation reduced the dimension of the integral from $6$ to $3$
and helped to find an easier numerical and analytical expression. 
We note here that we obtain a difference of a factor $1/2$ compared to JK, 
a discrepancy pointed out by them when comparing to previous work.

For illustration we show plots of the linear dominant 
contribution of the OV power spectrum in figure~(\ref{fig:pow_spect}) 
for the fiducial $\Lambda$CDM model assuming $z_r=8$ and 
$z_r=17$. The correspondence between 
full-sky multipole moments $C_{\ell}$ and the flat-sky Fourier space $P(K)$ is
straightforward $C_{\ell} = P(K=\ell)$. As expected, the $z_r=17$ 
scenario (keeping $x_e=1$) 
increases the amplitude of the power spectrum, 
due to the rise of the optical depth,
and shifts its peak towards smaller angular scales.
Numerically, we find for the amplitude of the power spectrum
the approximate scaling dependence 
$C_{\ell \simeq \, 500} \simeq 7.5\times10^{-18}\,x_e^2\,\log^{0.4}{(1+z_r)}$.
We advise the reader to consult JK paper~\cite{Jaffe:Kam98}
for the impact of changing the cosmological parameters (or
the reionization history) on the power and peak of the effect as well as for
the important detectability issues. The cosmological dependence
applies as well for the higher-order statistics. 
\subsubsection{Nonlinear extension: the kSZ effect power spectrum}
This extension was calculated previously~\cite{Hu99} and we present it 
for the dominant power spectrum for the sake of consistency.
The kinetic Sunyaev-Zel'dovich (kSZ) effect results from a Doppler 
effect suffered by the 
CMB photons as they travel through large scale structures 
emerged in a bulk flow. As pointed 
out by Hu~\cite{Hu99} among others~\cite{Ma:Fry01,Gnedin:Jaffe00}, in adiabatic 
CDM cosmologies, nonlinearities only affect the density field below the coherence 
scale of the bulk velocities and so the nonlinear density field is
uncorrelated with the large-scale velocity field, 
which remains linear. Hence, the mild 
nonlinear extension of the density contribution in the 
OV effect expressions naturally becomes 
the kSZ effect from large scale structures.

As the result of this, we can use the previous calculations 
of the OV effect power spectrum to obtain 
a similar expression for the kSZ by introducing a nonlinear 
correction in the density field contribution. Following Hu~\cite{Hu99} 
approach, we replace the linear density power spectrum with
its nonlinear analogue but leave the contribution from the velocity
power spectrum the same
\begin{equation}
P^{NL}_{dom}(K)=\frac{1}{8\pi^2}\int_{0}^{\eta_{0}}\frac{g^2(\eta)}{\eta^2}[a(\eta)]^2
                    \left( \frac{ {\mbox{\.{D}}}D}{D_0}\right)^2
                     \frac{P^{NL}(K/\eta)}{P(K/\eta)}S(K/\eta)d\eta
\label{eq:ovNL_effect_JK}
\end{equation}
where $P^{NL}$ stands for the nonlinear CDM power spectrum and where
the mode coupling integral $S$ was calculated previously  
(Eq.~(\ref{eq:S})) under linear theory. This expression includes both the
OV and the kSZ effects.
 
To calculate the nonlinear power spectrum, we assume that
the baryonic gas traces the dark matter~\cite{Hu99}. 
We follow Hamilton \emph{et. al.}~\cite{Hamilton91} who presented a 
scaling relation for the correlation function in the nonlinear 
regime that was generalized
to its Fourier analog by Peacock and Dodds~\cite{Pea:Dod94}. 
The basic hypothesis is that nonlinear fluctuations 
on a scale $k$ arise from linear
fluctuations on a larger scale
\begin{equation}
k_{\rm lin} = [1+\Delta_{\delta_c}^2(k)]^{-1/3} k 
\label{eq:kl_knl}
\end{equation}
so that there is a function relating the nonlinear and linear power
spectra at these two scales
\begin{equation}
\Delta_{\delta_c}^{2}(k) = f_{\rm NL}[\Delta_{\delta_c}^{2\, {\rm (lin)}}
	(k_{\rm lin})]
\label{eqn:PDscaling}
\end{equation}
which can be fit to simulations. We use the~\cite{Pea:Dod96} 
proposed expression for $f_{\rm NL}$. For stronger 
nonlinearities other corrections are necessary. 
See for example \cite{Zaldarriaga99}. One also needs the relation 
$\Delta_{\delta}^{2}(k)=k^3/(2\pi^2)P(k)$.

\begin{figure}[t]
\centerline{\epsfxsize=3.5truein\epsffile{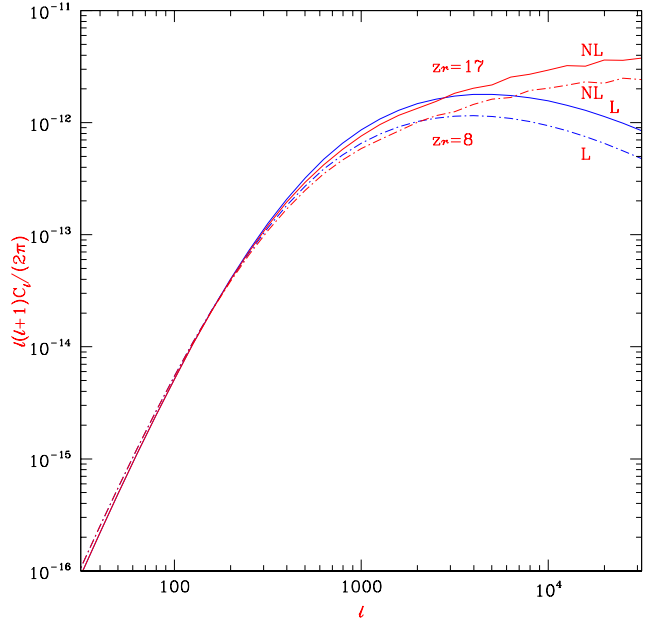}}
\caption{ \small{ The linear (label L) and nonlinear (label NL) OV power 
spectrum for the fiducial $\Lambda$CDM model. The {\it dot dash} 
lines correspond to $z_r=8$ and 
the {\it solid} lines to $z_r=17$. In both cases we assume
$\Delta z_r= 0.1(1+z_r)$ .}}
\label{fig:pow_spect}
\end{figure}

As discussed in W.Hu~\cite{Hu99}, this estimative
should be seen as an upper limit of the kSZ because on very 
small scales the gas pressure, unaccounted for, smooths 
the gas density as compared to the the dark matter 
density. The assumption that the baryonic gas 
traces the dark matter was shown to break down at multipoles 
$\ell\geq10^4$~\cite{Hu99,Gnedin:Hui98}.
We show a plot of the power spectrum corrected for mild 
nonlinearities in figure~(\ref{fig:pow_spect}) for illustration 
of the nonlinear enhancement of the power spectrum at small scales
for $z_r=8$ and $z_r=17$.
\section{The bispectrum of the OV effect}
\label{sec:Bispectrum}
\setcounter{subsubsection}{0}
\subsubsection{Linear bispectrum}
In analogy with the power spectrum, the flat-sky 
bispectrum of the OV effect 
is connected by Eq.~(\ref{eq:P-B-T}) to the following expectation value
\begin{eqnarray}
	<\frac{\tilde{\Delta T}}{\bar{T}}(\vec{K_1})
 	\frac{\tilde{\Delta T}}{\bar{T}}(\vec{K_2}) 
 	\frac{\tilde{\Delta T}}{\bar{T}}(\vec{K_3}) > 
               &=&    \frac{-1}{6}\int_{0}^{\eta_0}d\eta_1 g(\eta_1) 
                       \int_{0}^{\eta_0}d\eta_2 g(\eta_2)
                       \int_{0}^{\eta_0}d\eta_3 g(\eta_3) 
                       \int d^2\theta_1
                       \int d^2\theta_2
                       \int d^2\theta_3 \nonumber\\
               & &     \int_{}\frac{d^3q_1}{(2\pi)^3}
                       \int_{}\frac{d^3q_2}{(2\pi)^3}
                       \int_{}\frac{d^3q_3}{(2\pi)^3}\,
                       \hat{\theta}_{1i}
                       \hat{\theta}_{2j}
                       \hat{\theta}_{3l}\,
                       (<\tilde{p}_i({\mathbf q_1},\eta_1)
                        \tilde{p}_j({\mathbf q_2},\eta_2)
                        \tilde{p}_l({\mathbf q_3},\eta_3)> + \: perm.)\nonumber\\
               & &     e^{ i(\vec{K_1}.\vec{\theta}_1-\eta_1{\mathbf q_{1}}.\hat{\theta}_1) }
                       e^{ i(\vec{K_2}.\vec{\theta}_2-\eta_2{\mathbf q_{2}}.\hat{\theta}_2) }
                       e^{ i(\vec{K_3}.\vec{\theta}_3-\eta_3{\mathbf q_{3}}.\hat{\theta}_3) }
\label{eq:bispectrum}
\end{eqnarray}
where the five permutations are with respect to the ordering 
$({\mathbf q}_1,{\mathbf q}_2,{\mathbf q}_3)$.
Proceeding in a similar way we did for the derivation of the
power spectrum, this numerically heavy expression to integrate
can be considerably simplified by using a generalization of the Limber 
equation to higher-order statistics (see Appendix). 

We start by calculating a simplified expression for the first 
permutation $<\tilde{p}_i \tilde{p}_j \tilde{p}_l>$. 
At the end we generalize the 
results to include the total six permutations. 
Using expression~(\ref{eq:p}) 
for $\tilde{p}$ and the Wick theorem 
for the Gaussian 3-dimensional density field $6$-point correlation function 
$<\delta({\mathbf k}_1)\delta({{\mathbf q}}_1-{\mathbf k}_1)\delta({\mathbf k}_2)\delta({{\mathbf q}}_2-{\mathbf k}_2)\delta({\mathbf k}_3)\delta({{\mathbf q}}_3-{\mathbf k}_3)>$ 
we obtain $C^{6}_{2}.C^{4}_{2}/3!=15$ terms for 
$<\tilde{p}_i \tilde{p}_j \tilde{p}_l>$ of which $8$ are non-zero.
After some simple calculations, these $8$ terms can be condensed into
\begin{equation}
	<\tilde{p}_i({{\mathbf q}}_1,\eta_1)
         \tilde{p}_j({{\mathbf q}}_2,\eta_2)
         \tilde{p}_l({{\mathbf q}}_3,\eta_3)>=
          	       4\,G(\eta_1)G(\eta_2)G(\eta_3)
                       \,[F_{ijl}({{\mathbf q}}_1,{{\mathbf q}}_2)+
                        F_{ijl}({{\mathbf q}}_1,{{\mathbf q}}_3)]\,          
                       \delta^3_D({{\mathbf q}}_1+{{\mathbf q}}_2+{{\mathbf q}}_3)
\label{eq:p_p_p}
\end{equation}
with the time dependence function $G(\eta)=\left(\frac{iaD\dot{D}}{2D_0^2}\right)$ 
and the general tensorial functions $F_{\alpha\beta\gamma}$ given by
\begin{eqnarray}
F_{\alpha\beta\gamma}({{\mathbf q}}_i,{{\mathbf q}}_j) &=& \int{}{} d^3K'P(a)P(b)P(c)
             \left( \frac{a_{\alpha}}{a^2}-\frac{b_{\alpha}}{b^2} \right)
             \left( \frac{c_{\beta}}{c^2}-\frac{b_{\beta}}{b^2} \right)
             \left( \frac{a_{\gamma}}{a^2}-\frac{c_{\gamma}}{c^2} \right)
\label{eq:F_bis1}
\end{eqnarray}
where ${\mathbf a}={\mathbf K'}$, ${\mathbf b}={\mathbf K'}-{{\mathbf q}}_i$ and 
${\mathbf c}={\mathbf K'}+{{\mathbf q}}_j$. We concentrate on $F$ and combine
all the terms in Eq.~(\ref{eq:p_p_p}) when appropriate.\vspace{4mm}

Again using the small-angle approximation for which $\hat{\theta} \simeq (0,0,1)$, 
we can contract the $\hat{\theta}$'s with the vectors ${\mathbf a}$, 
${\mathbf b}$ and ${\mathbf c}$ such that we are left with the line of sight components 
of ${\mathbf a}$, ${\mathbf b}$ and ${\mathbf c}$. We can thus define 
a new scalar function $F$ such that 
\begin{eqnarray}
 F({{\mathbf q}}_i,{{\mathbf q}}_j)=
           \hat{\theta}_{1\alpha}\hat{\theta}_{2\beta}\hat{\theta}_{3\gamma}F_{\alpha\beta\gamma} &\simeq& 
             \int{}{} d^3K'P(a)P(b)P(c)
             \left( \frac{a_z}{a^2}-\frac{b_z}{b^2} \right)
             \left( \frac{c_z}{c^2}-\frac{b_z}{b^2} \right)
             \left( \frac{a_z}{a^2}-\frac{c_z}{c^2} \right) .
\label{eq:F_bis2}
\end{eqnarray}
Expanding this expression in different orders in $K_z'$ will give us 
the different levels of contributions under the Limber approximation 
to the bispectrum
\begin{eqnarray}
F({{\mathbf q}}_i,{{\mathbf q}}_j)
	             &=& \int{} d^3K'P(a)P(b)P(c) \nonumber\\
                     & & \left[ \right.
                          K_z'^3
                          \left( 
                           \frac{1}{a^2 b^4} -\frac{1}{a^4 b^2}-\frac{1}{a^2 c^4}+
                           \frac{1}{b^2 c^4} +\frac{1}{a^4 c^2}-\frac{1}{b^4 c^2}
                          \right) +\nonumber\\
                     & & K_z'^2
                          \left( 
                           q_{iz} \left( -\frac{2}{a^2b^4}+\frac{1}{a^4b^2}-\frac{1}{b^2c^4}
                                        +\frac{2}{b^4c^2}
                                  \right)+
                           q_{jz} \left(-\frac{2}{a^2c^4}+\frac{2}{b^2c^4}+\frac{1}{a^4c^2}
                                        -\frac{1}{b^4c^2} 
                                  \right)
                          \right) +\nonumber\\
		     & & K_z'
                          \left( 
	                   q_{iz}^2 \left( \frac{1}{a^2b^4}-\frac{1}{b^4c^2}
                                  \right)+
                           q_{jz}^2 \left( \frac{1}{b^2c^4}-\frac{1}{a^2c^4}
                                  \right)+
                           q_{iz}q_{jz}\left( \frac{2}{b^4c^2}-\frac{2}{c^4b^2}
                                  \right)
                         \right) +\nonumber\\
		     & &
                          \left( 
                           q_{iz}^2q_{jz} \left( -\frac{1}{b^4c^2}
                                          \right)+
                           q_{jz}^2q_{iz} \left( - \frac{1}{b^2c^4}
                                          \right)
                         \right)
                       \left. \right] .
\label{eq:F_bis3}
\end{eqnarray}
Again, and as detailed before in Sec.~\ref{subsec:PowSpectrum},
the integral in $K_z'$ for odd products of $K_z'$ is zero, 
and we are left with non-zero contributions 
from the second term (with $K_z'^2$) and 
the last term (with $K_z'^0$). The dominant contribution 
will come from the $K_z'^2$ though it 
has a dependency on $q_{z}$ which, by the Limber approximation, tends 
to be suppressed compared to terms which don't have such 
dependency. This result again confirms the discussion in Sec.~\ref{subsec:DomCont}, 
which predicts that the bispectrum of the OV effect has contributions only 
from terms with a dependency on $q_{iz}$ or on products of 
$3$ $q_{z}$.
   
We are interested on the dominant $K_z'^2$ term which can 
be written as
\begin{eqnarray}
F({{\mathbf q}}_i,{{\mathbf q}}_j)
	            &=& q_{iz}f_1({{\mathbf q}}_i,{{\mathbf q}}_j) + 
                        q_{jz}f_2({{\mathbf q}}_i,{{\mathbf q}}_j)
\label{eq:F_bis4}
\end{eqnarray}
where 
\begin{eqnarray}
f_1({{\mathbf q}}_i,{{\mathbf q}}_j)
	            &=& \int{} d^3K'P(a)P(b)P(c) 
                          K_z'^2
                          \left(   -\frac{2}{a^2b^4}+\frac{1}{a^4b^2}-\frac{1}{b^2c^4}
                                        +\frac{2}{b^4c^2}
                          \right)
\label{eq:F_bis5}
\end{eqnarray}
and
\begin{eqnarray}
f_2({{\mathbf q}}_i,{{\mathbf q}}_j)
	            &=& \int{} d^3K'P(a)P(b)P(c) 
                         K_z'^2
                          \left(   -\frac{2}{a^2c^4}+\frac{2}{b^2c^4}+\frac{1}{a^4c^2}
                                        -\frac{1}{b^4c^2} 
                          \right) .
\label{eq:F_bis6}
\end{eqnarray}

If we were to replace directly these last three expressions 
in~(\ref{eq:bispectrum})
and then apply the Limber approximation we would have 
a direct cancellation. We wish
to obtain an expression non-null which enhances the dominant 
character, as compared to the other sub-dominant terms, 
of this second-order contribution. 
To do so, we use again the small-angle approximation and consider 
$\hat{\theta}_i \simeq (0,0,1)$ to write
\begin{eqnarray}
F({{\mathbf q}}_i,{{\mathbf q}}_j)
	            &=& \hat{\theta}_i.q_{i} f_1({{\mathbf q}}_i,{{\mathbf q}}_j) + 
                        \hat{\theta}_j.q_{j} f_2({{\mathbf q}}_i,{{\mathbf q}}_j) .
\label{eq:F_bis7}
\end{eqnarray}
Now we replace this particular equation in (\ref{eq:p_p_p}) and then in 
(\ref{eq:bispectrum}) and finally integrate once 
by parts the time dependence. This is the
step that allows us to keep this term, which was expected to
be suppressed under the Limber approximation. 
Then we follow Kaiser's method and apply 
the Limber approximation, and using 
(\ref{eq:P-B-T}) we obtain the expression for one of the six possible
permutation terms of the bispectrum 
\begin{equation}
  	B^{OV}_{dom}(K_1,K_2,K_3) = \frac{i}{16\pi^3} \int_{0}^{\eta_{0}}d\eta \,
                        \frac{h^2}{\eta^4}\, \frac{\partial h}{\partial \eta}
                        \,\left[ f(\vec{q}_1,\vec{q}_2)+f(\vec{q}_1,\vec{q}_3) \right]
\label{eq:bispetrum_dom}
\end{equation}
where $h(\eta)=\left(g\frac{aD\dot{D}}{D_0^2}\right)$ and $f(\vec{q}_i,\vec{q}_j)=f_1(\vec{q}_i,\vec{q}_j)+f_2(\vec{q}_i,\vec{q}_j)$.

In order to simplify the functions $f_1$ and $f_2$, we need to find an explicit 
relation between ${\mathbf a}$, ${\mathbf b}$, ${\mathbf c}$ and 
${{\mathbf q}}_i$, ${{\mathbf q}}_j$ and ${\mathbf K'}$. 
To do so, we express both ${\mathbf K'}$ 
and ${\mathbf a}$, ${\mathbf b}$, ${\mathbf c}$ in the basis 
$(\hat{e}_z,\hat{q}_i,\hat{q}_j)$ where we remind the reader that 
$\hat{q}_{\alpha}\simeq(\hat{q}_{\alpha\perp},0)$ under the Limber 
approximation. Again, as for the power spectrum (see discussion 
in Sec.~\ref{subsec:PowSpectrum}), 
this suppresses any sub-dominant contribution that
could arise in the process but leaves our dominant contribution intact. 
We obtain
\begin{eqnarray}
	f({{\mathbf q}}_i,{{\mathbf q}}_j)
                     &=& f(\mid{{\mathbf q}}_i\mid,
                           \mid{{\mathbf q}}_j\mid,
                           \hat{q}_i.\hat{q}_j)\nonumber\\
                     &=& \int dK'_z \int du \int dv \sqrt{1-\mu^2}P(a)P(b)P(c) 
                         K_z'^2
                          \left(  \frac{1}{a^4b^2}+\frac{1}{a^4c^2}-\frac{2}{a^2b^4}
                                 -\frac{2}{a^2c^4}+\frac{1}{b^2c^4}+\frac{1}{b^4c^2}
                          \right)
\label{eq:F_bis8}
\end{eqnarray}
where $a=(K_z'^2+u^2+v^2+2uv\mu)^{1/2}$, 
$b=(K_z'^2+(u-q_i)^2+v^2+2(u-q_i)v\mu)^{1/2}$, 
$c=(K_z'^2+(v+q_j)^2+u^2+2(v+q_j)u\mu)^{1/2}$ and $\mu=\hat{q}_i.\hat{q}_j$.
As this expression corresponds to only one of the permutation terms of expression 
(\ref{eq:bispectrum}), we need to include the other $5$ to obtain the final expression 
for the dominant flat-sky bispectrum
\begin{equation}
	B^{OV}_{dom}(K_1,K_2,K_3) = \frac{i}{8\pi^3} \int_{0}^{\eta_{0}}d\eta\,
                         \frac{h^2}{\eta^4}\,\frac{\partial h}{\partial \eta}
                     	 \sum_{i,j=1;i\neq j}^{n=3} 
                         f \left( \frac{ \vec{K}_i }{\eta},
                                  \frac{ \vec{K}_j }{\eta}
                           \right)
\label{eq:bispetrum_subdom}
\end{equation}
where $f$ is given by (\ref{eq:F_bis8}).
We thus reduced our initial expression with twelve integrations to a 
four-dimensional integral, which can be numerically calculated for a 
chosen configuration of the wavenumbers $K$. 

Note that we obtain the first order time derivative of $h$. It
involves one single derivation, which indicates the $q_{iz}$ cancellation.
As $h$ is smooth and 
slowly varying, the contribution from this term should be very small.
Hence, the Limber cancellation at small scales reveals itself in
the derivatives of the time dependent functions.
\subsubsection{Nonlinear extension: the kSZ effect bispectrum}
We follow the same approach as for the power spectrum in Sec.~\ref{subsec:PowSpectrum}
and apply it to the dominant contribution.
As we have three CDM power spectrum showing up
in the expression for the bispectrum, which we need to divide among 
linear/nonlinear contributions from density/velocity contributions we will  
assume the following bispectrum effect
\begin{equation}
	B^{0V-NL}_{dom}(K_1,K_2,K_3) = \frac{i}{8\pi^3} \int_{0}^{\eta_{0}}d\eta\,
                         \frac{h^2}{\eta^4}\,\frac{\partial h}{\partial \eta}
                     	 \sum_{i,j,k=1;i\neq j}^{n=3}
                         \left( \frac{P^{NL}(K_k/\eta)}{P(K_k/\eta)} \right)^{3/2}
                         f \left( \frac{ \vec{K}_i }{\eta},
                                  \frac{ \vec{K}_j }{\eta}
                           \right) .
\label{eq:bispetrumNL_dom}
\end{equation}
\subsubsection {The Signal-to-noise}
\label{subsec:S_N_bispectrum}
To define the $\chi^2$ statistics of Eq.~(\ref{eq:chi}), 
we need to calculate the likelihood 
$L$ of observing the bispectrum elements 
$B_{\beta} \equiv B_{\ell_1 \ell_2 \ell_3}$ given the true parameters ${\mathbf p}$, 
and calculate the Fisher matrix as defined in Eq.~(\ref{eq:fisher})
\begin{equation}
        F_{ij} \equiv - 
               \left<  
                     \frac{\partial^{2}ln L({\mathbf B};{\mathbf p})}
                          {\partial p_i \partial p_j} 
               \right> .
\label{eq:chi_bisp1}
\end{equation}
Assuming that the likelihood is Gaussian, we follow Cooray and Hu 
\cite{Cooray:Hu00} approach to calculate the $\chi^2$ statistics 
\begin{equation}
\chi^2 \equiv  \left( \frac{S}{N} \right)^2
          =    f_{sky}\sum_{\ell_3 \geq \ell_2 \geq \ell_1}
               \frac{ B^2_{\ell_1 \ell_2 \ell_3} }{\sigma^2_{\ell_1 \ell_2 \ell_3}},
\label{eq:chi_bisp2}
\end{equation}
where $B_{\ell_1 \ell_2 \ell_3}$ is the angular averaged bispectrum defined 
on the sphere, $f_{sky}$ represents the reduction in signal-to-noise
due to incomplete sky coverage and $\sigma^2_{\ell_1 \ell_2 \ell_3}$
comes from the covariance matrix of the angular averaged bispectrum assuming 
a nearly Gaussian bispectrum and full-sky 
coverage~\cite{Goldberg:Spergel99,Gangui:Martin00}
\begin{equation}
\sigma^2_{\ell_1 \ell_2 \ell_3}= C_{\ell_1}^{tot}C_{\ell_2}^{tot}C_{\ell_3}^{tot}\,
                [1 + \delta_D^3({\mathbf \ell_1}+{\mathbf \ell_2})+\delta_D^3({\mathbf \ell_2}+{\mathbf \ell_3})+\delta_D^3({\mathbf \ell_3}+{\mathbf \ell_1}) 
                   + 2\delta_D^3({\mathbf \ell_1}+{\mathbf \ell_2})\delta_D^3({\mathbf \ell_1}+{\mathbf \ell_3})]
\label{eq:sigma_bisp}
\end{equation}
where $C_{\ell}^{tot}$ stands for the sum of the power spectra of the 
primary cosmic signal, the thermal SZ (thSZ) effect which contributes
significantly at the scales of interest, the linear OV effect, the detector 
noise and the foregrounds respectively
\begin{equation}
C_{\ell}^{tot} = C_{\ell} + C_{\ell}^{thSZ} + C_{\ell}^{OV-L} + C_{\ell}^{noise} 
               + C_{\ell}^{foregrounds} .
\label{eq:cl_tot}
\end{equation}
The thermal SZ effect was taken from~\cite{Aghanim02}
and was calculated semi-analytically 
at $30$ GHZ for a normalization factor $\sigma_8=0.9$. We
include the linear OV effect contribution (see Sec.~\ref{subsec:PowSpectrum}) 
although its amplitude is small as
compared with the primary and the thermal SZ signals 
at the scales considered in this work. 
The primary cosmic signal was computed with CMBFAST~\cite{Seljak:Zaldarriaga96} 
and we will not consider any foreground. Indeed, for the case
of MAP and Planck, studies indicate that the total $C_{\ell}$
should increase by $10\%$ maximum~\cite{Tegmark99}. However, 
caution is required as this result assumes that foregrounds
have a Gaussian distribution. The foregrounds contribution 
to the higher-order statistics could in fact be our main 
obstacle in measuring non-Gaussian effects and very little 
is known about them. For the noise power spectrum 
we use Eqs.~(\ref{eq:cl_noise}--\ref{eq:cl_noise_freq}) 
applied to MAP and Planck. 

We will study the contributions to the 
$\chi^2$ per log interval in $\ell$. It gives us more sensibility
on the angular scales of stronger detectability, depending
of the effects considered. This also enables us to directly 
compare our results with Cooray and Hu~\cite{Cooray:Hu00}.
We calculate the total $S/N$ for each experiment by
integrating over the multipole $\ell$.

A useful relation between the flat-sky bispectrum and the spherical
harmonic angular averaged bispectrum~\cite{Komatsu:Spergel00} is
\begin{equation}
B_{\ell_1 \ell_2 \ell_3}= \sqrt{ \frac{(2\ell_1+1)(2\ell_2+1)(2\ell_3+1)}{4\pi} }
	           \left( \begin{array}{ccc}
                              \ell_1 & \ell_2 &\ell_3 \\
                              0  &  0  & 0
                           \end{array}
                   \right)
                  B(K_1=\ell_1,K_2=\ell_2,K_3=\ell_3) .
\label{eq:bisp_flat_sph}
\end{equation}
Since the Wigner-3j vanishes if $\ell_1+\ell_2+\ell_3=odd$, the full-sky 
bispectrum can only be estimated for even terms.
\subsubsection{Results}
We are concerned with the overall detectability of the dominant term,
previously calculated. 
Hence we choose the simplest of the possible configurations,
highly localized in Fourier space, $K_1=K_2=K_3=l$, for which 
the flat-sky dominant bispectrum becomes
\begin{equation}
	B^{OV}_{dom}(K=\ell) = \frac{3i}{4\pi^3} \int_{0}^{\eta_{0}}d\eta\,
                         \frac{h^2}{\eta^4}\,\frac{\partial h}{\partial \eta}\,
                         f \left( \frac{\ell}{\eta} \right)
\label{eq:bispetrum_dom_red}
\end{equation}
where $f$ is defined by Eq.~(\ref{eq:F_bis8}). The nonlinear 
analog follows from the previous equation.

With this configuration, depending only on the multipole $\ell$, 
there is a simple way of calculating an estimate 
of the order of magnitude of the $(S/N)^2$ per bin of $\ell$
were we to include all the remaining $\ell$ 
modes~\cite{Komatsu:Spergel00} of the full-sky bispectrum. 
Indeed in the Eq.~(\ref{eq:chi_bisp2}), 
we see that the number of modes contributing to the $(S/N)^2$ 
per bin of $\ell$ increases
as $\ell^{2}$ and in Eq.~(\ref{eq:bisp_flat_sph}) $\ell^3
           \left( 
               \begin{array}{ccc} 
                     \ell & \ell & \ell \\ 
                     0 & 0 & 0 
                \end{array} \right)^2
                  $ increases as $0.36\,\ell$ so
\begin{equation}
\frac{d\chi^2}{d\ell}  \sim f_{sky}\,\ell^{2}\,\frac{B_{\ell}^2}{\sigma^2_{\ell}}
        \sim f_{sky}\,\ell^{2}\,
           \ell^3\left( 
            \begin{array}{ccc} \ell & \ell & \ell \\ 
                               0 & 0 & 0 
            \end{array} 
                \right)^2
           \frac{B^2(\ell)}{\sigma^2_{\ell}}
        \sim  0.36\,f_{sky}\,\ell^{3}\,\frac{B^2(K=\ell)}{\sigma^2_{\ell}} .
\label{eq:chi_bisp_est}
\end{equation}
The $\sigma^2_{\ell}$ is calculated using Eq.~(\ref{eq:sigma_bisp}) 
for all the $\ell$s equal and $B(\ell)$ using Eq.~(\ref{eq:bispetrum_dom_red}).
\vspace{4mm}

\begin{figure*}
\centering 
\vbox to 0.30\textheight{
{\hbox to \textwidth{
\null\null
\epsfxsize=0.45\textwidth
\epsfysize=0.30\textheight
\epsfbox{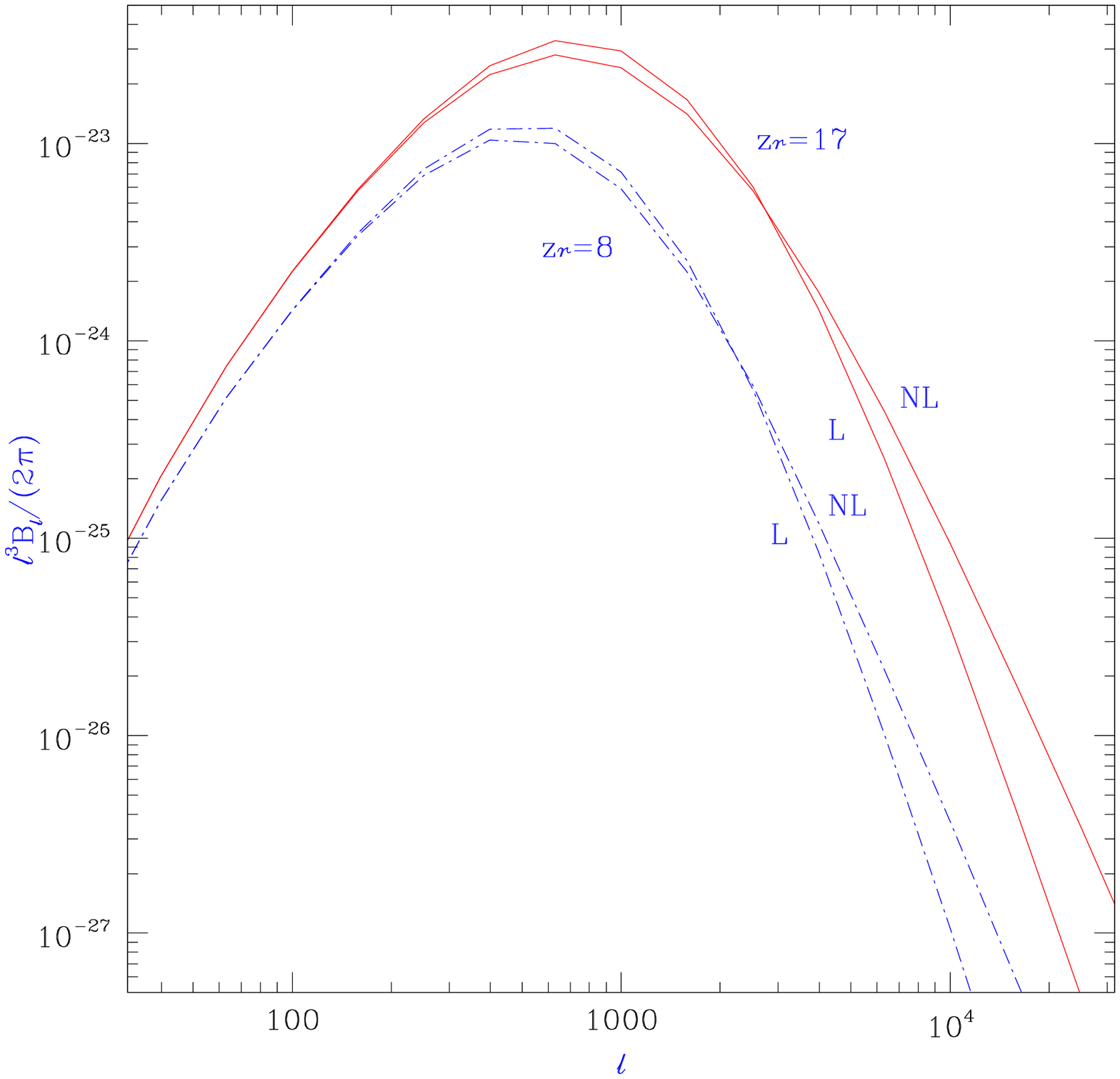}
\hfill
\epsfxsize=0.45\textwidth
\epsfysize=0.30\textheight
\epsfbox{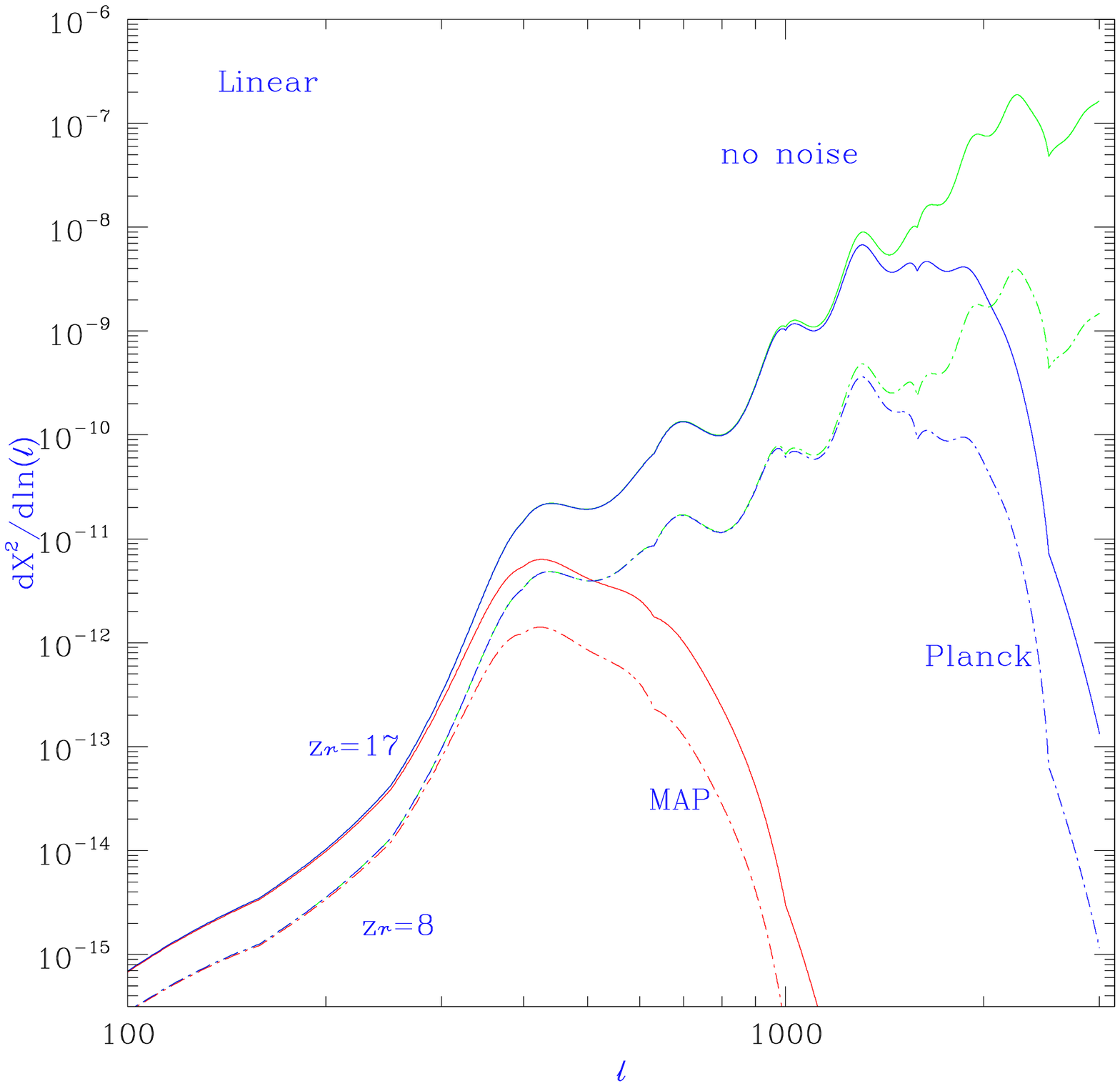}
\hfill
}}}
\caption{ \small{ 
{\it Left panel---} Linear (label L) flat-sky bispectrum of the OV effect
and its nonlinear extension (label NL).
{\it Right panel ---} Contribution to $\chi^2$ per log interval 
in $\ell$ for the OV full-sky linear bispectrum with no instrumental noise 
({\it top}), Planck noise ({\it middle}) and MAP noise ({\it bottom}) included 
in the variance. We used the specifications in the table~\ref{table1}.
All the plots were calculated for the fiducial $\Lambda$CDM model. The 
{\it dot dash} lines correspond to $z_r=8$ and 
the {\it solid} lines to $z_r=17$. We assumed
$\Delta z_r= 0.1(1+z_r)$.
The total $S/N$ for the OV full-sky linear bispectrum for MAP, Planck 
and a perfect experiment respectively are: 
$5.0\times10^{-5}$, $1.2\times10^{-3}$ and $4.4\times10^{-3}$ assuming $z_r=8$,
and $3.7\times10^{-5}$, $2.0\times10^{-3}$ and $1.4\times10^{-2}$ assuming $z_r=17$.
}}
\label{fig:bispectrum}
\null
\end{figure*}

We show the plots for the linear and the nonlinear flat-sky OV bispectrum in 
figure~(\ref{fig:bispectrum}) for $z_r=8$ and $z_r=17$ in the context 
of our fiducial cosmological model. For $z_r=8$ the peak of the effect 
occurs around multipole $l \simeq 500$ with an amplitude of 
$l^3B(\ell)/(2\pi) \simeq 1.2\times10^{-23}$
whereas for $z_r=17$ the peak takes place at a higher multipole
of $\ell \simeq 700$ with a stronger amplitude of 
$\ell^3B(\ell)/(2\pi) \simeq 3\times10^{-23}$, as expected. 
We find numerically for the amplitude of the bispectrum
the approximate scaling dependence with reionization history
$B(\ell \simeq \,50) \simeq 1.7\times10^{-29}\,x_e^3\,\log{(1+z_r)}$. The
dependence with $z_r$ is neither in agreement with the one obtained 
for the power spectrum (see Sec.~\ref{subsec:PowSpectrum}) 
nor with the one obtained for the trispectrum (see Sec.~\ref{sec:Trispectrum}).
This can be explained by looking at the bispectrum 
Eq.~(\ref{eq:bispetrum_dom_red}) where we have 
the first order time derivative of $h$, contrary to
the corresponding expressions for the power spectrum 
(Eq.~(\ref{eq:ov_effect_JK}))
and the trispectrum (Eq.~(\ref{eq:trispectrum_dom_red})).
This derivative of $h$ introduces a stronger scaling 
relation with $z_r$ for the amplitude of the bispectrum
as compared to the one for the other two statistics.
For both reionization scenarios,
the most interesting feature is the rapid drop of power after the peak 
which is not observed for the power spectrum 
(see figure~(\ref{fig:pow_spect})), which is a 
direct consequence of the Limber cancellation at small angular scales. 
The fact that the bispectrum is not considerable enhanced 
by nonlinearities, which take place at small angular scales, is also
the result of the effect peaking at intermediate multipoles. The second,
but expected, result is the low overall amplitude of the effect.
The $z_r=8$ case (which we can easily compare with previous results in the 
literature) is lower by more than $10$ orders of magnitude
than the lensing couplings presented by Cooray and Hu~\cite{Cooray:Hu00}, 
and by a few orders of magnitude than most of the OV couplings
involving the SZ effect 
though it should be comparable to the OV
couplings involving the Doppler or ISW/SW effects.
Despite the fact that such OV couplings 
suffer the Limber cancellation 
as well (they correspond to expectation
values of an odd product of a vector field), 
this cancellation can be counterbalanced
by the coupling to higher amplitude effects,
like the SZ effect, and by the matching in 
redshift between the density and velocity
fields of the OV effect and the secondaries to which is couples. 
This is the case of the hybrid coupling ISW-SZ-OV 
presented by~\cite{Cooray:Hu00}, 
that has the largest signal of all secondaries
that couple with OV. 

A more revealing quantity than the power of the effect 
is the signal-to-noise ratio. We plot in figure~(\ref{fig:bispectrum}),
the estimated contributions to $\chi^2$ 
per log interval in $\ell$ of the linear full-sky bispectrum 
had we included all the modes for MAP, Planck~\cite{MAP/PLANCK} 
and no instrumental noise. We show the results for $z_r=8$ and $z_r=17$. 
The experimental specifications can be found in table~\ref{table1}. 
We found no necessity of plotting the corresponding 
$d \chi^2 / d\ell$ for the nonlinear bispectra
due to its similarity to the linear 
bispectra. The structure in the $d\chi^2/d\ell$
arises mainly from the structure of the CMB primary
power spectrum at $\ell > 200$. We point out that the continuous rise 
at $\ell\simeq 3000$ for a perfect experiment is due to the fact that
the signal decreases slower than the contributions to the $C_{\ell}$
from primary, linear OV and thermal SZ anisotropies up to these scales.
Though it is not explicit in the figure, the 
$d\chi^2/d\ell$ is zero when $3\ell$ is odd.
As we can see, considering thermal SZ contributions 
to the noise evaluated at $30$ GHz, the signal-to-noise
of the OV bispectrum is very small, even for a perfect
experiment, for which the total $S/N \sim 4.4\times10^{-3}$
for $z_r=8$ and $S/N \sim 1.4\times10^{-2}$ for $z_r=17$. 
These values can be compared to the value $S/N \sim 1.7$ obtained by 
Cooray and Hu~\cite{Cooray:Hu00}
for the bispectrum generated by the coupling 
ISW-SZ$^{NL}$-OV.
Hence, even for early reionization,
no detection is to be expected from future
experiments unable to remove the thermal SZ effect.
For the ideal case of a perfect multi-frequency experiment
observing in the millimeter and sub-millimeter
capable of subtracting all of the thermal SZ effect, 
the total signal-to-noise increases
to $S/N \sim 0.1$ for $z_r=17$, which means that
not even for this case are we to expect a detection.
Our results for the $S/N$ rely on various convenient assumptions
required to considerably simplify the analytical expression 
for the estimate of the $S/N$
(see sections Sec.~\ref{subsec:Signal-to-noise} and 
Sec.~\ref{subsec:S_N_bispectrum}). This means that the values
obtained for the $S/N$ are most probably upper limits on the real 
values expected and should thus be considered accordingly.
\section{The trispectrum of the OV effect}
\label{sec:Trispectrum}
\setcounter{subsubsection}{0}
\subsubsection{Linear trispectrum}
We follow the same procedure used for the power spectrum and
the bispectrum. The trispectrum 
contribution to the temperature fluctuations is connected by Eq.
~(\ref{eq:P-B-T}) to
\begin{eqnarray}
\lefteqn{
 <\frac{\tilde{\Delta T}}{\bar{T}}(\vec{K_1})
 \frac{\tilde{\Delta T}}{\bar{T}}(\vec{K_2})
 \frac{\tilde{\Delta T}}{\bar{T}}(\vec{K_3})
 \frac{\tilde{\Delta T}}{\bar{T}}(\vec{K_4}) > =}\nonumber\\
               & &  \frac{1}{24}\int_{0}^{\eta_0}d\eta_1 g(\eta_1) 
                    \int_{0}^{\eta_0}d\eta_2 g(\eta_2)
                    \int_{0}^{\eta_0}d\eta_3 g(\eta_3) 
                    \int_{0}^{\eta_0}d\eta_4 g(\eta_4)
                    \int d^2\theta_1
                    \int d^2\theta_2
                    \int d^2\theta_3
                    \int d^2\theta_4 \nonumber\\
               & &\int_{}\frac{d^3q_1}{(2\pi)^3}
                  \int_{}\frac{d^3q_2}{(2\pi)^3}
                  \int_{}\frac{d^3q_3}{(2\pi)^3}
                  \int_{}\frac{d^3q_4}{(2\pi)^3}\,
                  \hat{\theta}_{1i}
                  \hat{\theta}_{2j}
                  \hat{\theta}_{3l}
                  \hat{\theta}_{4m}\,
             (<\tilde{p}_i({\mathbf q_1},\eta_1)
              \tilde{p}_j({\mathbf q_2},\eta_2)
              \tilde{p}_l({\mathbf q_3},\eta_3)
              \tilde{p}_m({\mathbf q_4},\eta_4)> +\:perm.)\nonumber\\
               & & e^{ i(\vec{K_1}.\vec{\theta}_1-\eta_1{\mathbf q_{1}}.\hat{\theta}_1) }
                   e^{ i(\vec{K_2}.\vec{\theta}_2-\eta_2{\mathbf q_{2}}.\hat{\theta}_2) }
                   e^{ i(\vec{K_3}.\vec{\theta}_3-\eta_3{\mathbf q_{3}}.\hat{\theta}_3) }
                   e^{ i(\vec{K_4}.\vec{\theta}_4-\eta_4{\mathbf q_{4}}.\hat{\theta}_4) }.
\label{eq:trispectrum}
\end{eqnarray}
The 24 total permutations arise from symmetries under permutation invariance.
Again we concentrate on the first of the permutations and then generalize at the end.
We obtain this time for the first permutation term of the previous Eq.
$<\tilde{p}_i \tilde{p}_j \tilde{p}_l \tilde{p}_m>$ 
by the Wick theorem applied to the Gaussian 3-dimensional 
density field $8$-point correlation function 
$C^{8}_{2}.C^{6}_{2}.C^{4}_{2}/4!=105$ terms of which
$12$ non-zero terms for the Gaussian contribution and $48$ 
non-zero terms for the connected part which interests us here. Performing
some exhaustive and systematic calculations these 
$48$ terms can be written in the following condensed form
\begin{eqnarray}
<\tilde{p}_i({\mathbf q_1},\eta_1)
              \tilde{p}_j({\mathbf q_2},\eta_2)
              \tilde{p}_l({\mathbf q_3},\eta_3)
              \tilde{p}_m({\mathbf q_4},\eta_4)> &=&
                     4\,G(\eta_1)G(\eta_2)G(\eta_3)G(\eta_4) \nonumber\\
                & & [F_{imlj}({{\mathbf q}}_1,{{\mathbf q}}_2,{{\mathbf q}}_3,{{\mathbf q}}_4)+
                     F_{ijlm}({{\mathbf q}}_1,{{\mathbf q}}_4,{{\mathbf q}}_2,{{\mathbf q}}_3)+
                     F_{lijm}({{\mathbf q}}_3,{{\mathbf q}}_4,{{\mathbf q}}_1,{{\mathbf q}}_2)+\nonumber\\
                & &  F_{ljmi}({{\mathbf q}}_3,{{\mathbf q}}_1,{{\mathbf q}}_2,{{\mathbf q}}_4)+ 
                     F_{jmli}({{\mathbf q}}_2,{{\mathbf q}}_3,{{\mathbf q}}_1,{{\mathbf q}}_4)+
                     F_{jilm}({{\mathbf q}}_2,{{\mathbf q}}_4,{{\mathbf q}}_1,{{\mathbf q}}_3)]\nonumber\\
                & &  \delta^3({{\mathbf q}}_1+{{\mathbf q}}_2+{{\mathbf q}}_3+{{\mathbf q}}_4) .
\label{eq:p_p_p_p}
\end{eqnarray}
where $G$ is defined in the previous section and the general tensorial 
function $F_{\alpha\beta\gamma\delta}$ is defined as
\begin{eqnarray}
F_{\alpha\beta\gamma\delta}({{\mathbf q}}_i,{{\mathbf q}}_j,{{\mathbf q}}_l,{{\mathbf q}}_m) &=&
          f_{\alpha\beta\gamma\delta}({{\mathbf q}}_i,{{\mathbf q}}_j,{{\mathbf q}}_l)+
          f_{\alpha\gamma\beta\delta}({{\mathbf q}}_i,{{\mathbf q}}_j,{{\mathbf q}}_m)
\label{eq:F_tris1}
\end{eqnarray}
where a general $f_{xyzw}$ is given by
\begin{eqnarray}
f_{xyzw}({{\mathbf q}}_i,{{\mathbf q}}_j,{{\mathbf q}}_l) &=&
           \int{} d^3K'P(a)P(b)P(c)P(d)
             \left( \frac{a_{x}}{a^2}+\frac{b_{x}}{b^2} \right)
             \left( \frac{d_{y}}{d^2}-\frac{c_{y}}{c^2} \right)
             \left( \frac{b_{z}}{b^2}+\frac{d_{z}}{d^2} \right)
             \left( \frac{a_{w}}{a^2}-\frac{c_{w}}{c^2} \right)
\label{eq:f_tris2}
\end{eqnarray}
where ${\mathbf a}={\mathbf K'}$, ${\mathbf b}={\mathbf K'}-{{\mathbf q}}_i$, 
${\mathbf c}={\mathbf K'}+{{\mathbf q}}_j$ and 
${\mathbf d}={\mathbf K'}+{{\mathbf q}}_j+{{\mathbf q}}_l$. Again we will work 
with a single $f$ and at the end generalize the result. \vspace{4mm}

Applying the small-angle approximation again we can remove the tensorial dependence 
of our expression and contract the $\theta_i$ with our vectors. 
We obtain the scalar $f$
\begin{eqnarray}
f({{\mathbf q}}_i,{{\mathbf q}}_j,{{\mathbf q}}_l) &=&
               \hat{\theta}_{1\alpha}\hat{\theta}_{2\beta}
               \hat{\theta}_{3\gamma}\hat{\theta}_{4\delta}
              f_{\alpha\beta\gamma\delta}
                   ({{\mathbf q}}_i,{{\mathbf q}}_j,{{\mathbf q}}_l) \nonumber \\
          &\simeq& \int{} d^3K'P(a)P(b)P(c)P(d)
             \left( \frac{a_z}{a^2}+\frac{b_z}{b^2} \right)
             \left( \frac{d_z}{d^2}-\frac{c_z}{c^2} \right)
             \left( \frac{b_z}{b^2}+\frac{d_z}{d^2} \right)
             \left( \frac{a_z}{a^2}-\frac{c_z}{c^2} \right)
\label{eq:f_tris3}
\end{eqnarray}
and the scalar $F$
\begin{eqnarray}
F({{\mathbf q}}_i,{{\mathbf q}}_j,{{\mathbf q}}_l,{{\mathbf q}}_m) &=&
          f({{\mathbf q}}_i,{{\mathbf q}}_j,{{\mathbf q}}_l) + f({{\mathbf q}}_i,{{\mathbf q}}_j,{{\mathbf q}}_m) .
\label{eq:F_tris4}
\end{eqnarray}
Expanding $f$ in order of $K_z'$ and keeping the non-nulls terms (see previous
section), i.e. the
terms which are even in $K_z'$, we are left with
\begin{eqnarray}
	f({{\mathbf q}}_i,{{\mathbf q}}_j,{{\mathbf q}}_l) &=&
                   \int{} d^3KP(a)P(b)P(c)P(d)\nonumber\\
               & &  \left[ \right.
                      K_z'^4
                          \left( \frac{1}{a^2}-\frac{1}{b^2} \right)
                          \left( \frac{1}{c^2}-\frac{1}{d^2} \right)
                          \left( \frac{1}{d^2}-\frac{1}{b^2} \right)
                          \left( \frac{1}{c^2}-\frac{1}{a^2} \right)\nonumber\\
                & & + K_z'^2
                          \left( \right. 
                           q_{iz}^2 \left( \right.\frac{1}{b^4c^4}-\frac{1}{a^2b^4c^2}
                                           +\frac{1}{a^4d^4}
                                           -\frac{1}{a^2b^2d^4}-\frac{1}{a^2c^2d^4}
                                           +\frac{1}{b^2c^2d^4}-\frac{1}{a^2b^4d^2}\nonumber\\
                & &                        -\frac{1}{a^4b^2d^2}-\frac{1}{b^2c^4d^2} 
                                           +\frac{1}{b^4c^2d^2}+\frac{2}{a^2b^2c^2d^2}
                                    \left. \right)\nonumber\\
                & &        +q_{jz}^2 \left( \frac{1}{b^4c^4}+\frac{1}{a^2b^2c^4}
                                          +\frac{1}{a^2c^4d^2}+\frac{1}{b^2c^4d^2}
                                    \right)
                           +q_{lz}^2 \left( \frac{1}{a^4d^4}+\frac{1}{a^2b^2d^4}
                                          -\frac{1}{b^2c^2d^4}-\frac{1}{b^2c^4d^2}
                                    \right)\nonumber\\
                & &        +q_{iz}q_{jz} \left( -\frac{4}{b^4c^4}-\frac{2}{a^2b^2c^4}
                                            +\frac{2}{a^2b^4c^2}+\frac{1}{a^4b^2c^2}
                                             -\frac{2}{a^2c^2c^4}-\frac{1}{b^2c^2d^4}
                                             +\frac{2}{a^2c^4d^2}-\frac{1}{a^4c^2d^2}
                                        \right)\nonumber\\
                & &        +q_{iz}q_{lz} \left( \frac{2}{a^4d^4}-\frac{2}{a^2c^2d^4}
                                          -\frac{2}{a^2b^4d^2}-\frac{1}{a^4b^2d^2}
                                          -\frac{1}{b^2c^4d^2}+\frac{2}{b^4c^2d^2}
                                          +\frac{2}{a^2b^2c^2d^2}
                                        \right)\nonumber\\
                & &        +q_{jz}q_{lz} \left( -\frac{2}{a^2c^2d^4}-\frac{2}{b^2c^2d^4}
                                          +\frac{2}{a^2c^4d^2}+\frac{2}{b^2c^4d^2}
                                          -\frac{1}{a^4c^2d^2}-\frac{1}{b^4c^2d^2}
                                          -\frac{2}{a^2b^2c^2d^2}
                                        \right)
                          \left. \right)+ \nonumber\\
		& &      \left( \right.
                           q_{iz}^3q_{jz} \left( \frac{1}{b^2c^2d^2}-\frac{1}{b^4c^2d^2}
                                          \right)
                           +q_{iz}^2q_{iz}^2 \left( \frac{1}{b^4c^2}-\frac{1}{b^2c^4d^2}
                                          \right)\nonumber\\
                & &        +q_{iz}^2q_{jz}q_{lz}\left( \frac{2}{b^2c^2d^4}-\frac{1}{b^4c^2d^2}
                                          \right)
 	                   +q_{iz}q_{jz}^2q_{lz}\left( -\frac{1}{b^2c^4d^2}
                                          \right)
                           +q_{iz}q_{jz}q_{lz}^2\left( \frac{1}{b^2c^2d^4}
                                          \right)
                         \left. \right)
                   \left. \right] .
\label{eq:f_tris5}
\end{eqnarray}
We concentrate on the dominant term which can be simplified
using the same method applied to the calculation of the dominant 
term of the power spectrum. It can be written as
\begin{eqnarray}
	f({{\mathbf q}}_i,{{\mathbf q}}_j,{{\mathbf q}}_l) &=&
                    \int{} d^3K'P(a)P(b)P(c)P(d)
                    K_z'^4
                    \left( \frac{1}{a^2}-\frac{1}{b^2} \right)
                    \left( \frac{1}{c^2}-\frac{1}{d^2} \right)
                    \left( \frac{1}{d^2}-\frac{1}{b^2} \right)
                    \left( \frac{1}{c^2}-\frac{1}{a^2} \right) .
\label{eq:f_tris6}
\end{eqnarray}
To find an explicit relation between ${\mathbf a}$, ${\mathbf b}$, 
${\mathbf c}$, ${\mathbf d}$ and ${{\mathbf q}}_i$, ${{\mathbf q}}_j$, 
${{\mathbf q}}_l$ and ${\mathbf K'}$ we express both ${\mathbf K'}$ and 
${\mathbf a}$, ${\mathbf b}$, ${\mathbf c}$, ${\mathbf d}$ in the basis 
$(\hat{e}_z,\hat{q}_i,\hat{q}_j)$. We should say here that using the Limber 
approximation will allow only two of our initial vectors among the set of four 
(${{\mathbf q}}_1$,${{\mathbf q}}_2$,${{\mathbf q}}_3$,${{\mathbf q}}_4$) to be 
independent. Indeed all parallel components to the line of sight will 
be negligible and the four vectors will be inside the same plane, the 
one perpendicular to the line of sight. This justifies the use of the 
basis chosen. We stress one time more that this eliminates any
sub-dominant term that could show up (see discussion 
in Sec.~\ref{subsec:PowSpectrum}). 
So in that basis our function $f$ can be expressed in 
the following simplified way
\begin{eqnarray}
	f({{\mathbf q}}_i,{{\mathbf q}}_j,{{\mathbf q}}_l)&=&
                  f(\mid {{\mathbf q}}_i \mid,\mid {{\mathbf q}}_j \mid,\mid {{\mathbf q}}_l \mid,
                    \hat{q}_i.\hat{q}_j,\hat{q}_i.\hat{q}_l,\hat{q}_j.\hat{q}_l)
                     \nonumber\\
              &=&   \int{}dK'_z \int{}du \int{}dv 
                    \sqrt{1-\alpha^2}P(a)P(b)P(c)P(d)K_z'^4
                    \left( \frac{1}{a^2}-\frac{1}{b^2} \right)
                    \left( \frac{1}{c^2}-\frac{1}{d^2} \right)
                    \left( \frac{1}{d^2}-\frac{1}{b^2} \right)
                    \left( \frac{1}{c^2}-\frac{1}{a^2} \right)
\label{eq:f_tris7}
\end{eqnarray}
where $\alpha=\hat{q}_i.\hat{q}_j$, $\beta=\hat{q}_i.\hat{q}_l$ 
and $\gamma=\hat{q}_j.\hat{q}_l$. Also $a=(K_z'^2+u^2+v^2+2uv\alpha)^{1/2}$, 
$b=(K_z'^2+(u-q_i)^2+v^2+2(u-q_i)v\alpha)^{1/2}$, 
$c=(K_z'^2+(v+q_j)^2+u^2+2(v+q_j)u\alpha)^{1/2}$, 
$d=(K_z'^2+(v+q_j+y)^2+(u+x)^2+2(v+q_j+y)(u+x)\alpha)^{1/2}$ with 
$y=q_l\frac{\gamma-\beta\alpha}{(1-\alpha^2)}$ and 
$x=q_l\frac{\beta-\gamma\alpha}{(1-\alpha^2)}$. Our $f$ only depends 
on the norms of its arguments and on the angles between their directions.
We can combine these expressions to simplify the expression 
(\ref{eq:trispectrum}). Following Kaiser's method and proceeding 
with the Limber approximation, we obtain the expression for 
one of the $24$ possible permutation terms of the trispectrum
\begin{eqnarray}
\lefteqn{
  	T^{OV}_{dom}(K_1,K_2,K_3,K_4) =
             \frac{1}{32\pi^3} \int_{0}^{\eta_{0}}d\eta
           \left( \frac{ aD\dot{D}}{ D_0^2 } \right)^4 
             \frac{g(\eta)^4}{\eta^6} }\nonumber\\
           & & [\,F(\vec{K}_1/\eta,\vec{K}_2/\eta,\vec{K}_3,/\eta\vec{K}_4/\eta)+
                F(\vec{K}_1/\eta,\vec{K}_4/\eta,\vec{K}_2/\eta,\vec{K}_3/\eta)+
                F(\vec{K}_3/\eta,\vec{K}_4/\eta,\vec{K}_1/\eta,\vec{K}_2/\eta)+\nonumber\\
           & &  F(\vec{K}_3/\eta,\vec{K}_1/\eta,\vec{K}_2/\eta,\vec{K}_4/\eta)+ 
                F(\vec{K}_2/\eta,\vec{K}_3/\eta,\vec{K}_1/\eta,\vec{K}_4/\eta)+
                F(\vec{K}_2/\eta,\vec{K}_4/\eta,\vec{K}_1/\eta,\vec{K}_3/\eta)\,]
\label{eq:trispectrum_dom}
\end{eqnarray}
where $F$ is defined by Eq.~(\ref{eq:F_tris4}) and the $f$ by 
Eq.~(\ref{eq:f_tris7}).
Finally, by including all the permutation terms in Eq.~(\ref{eq:trispectrum}) we obtain
\begin{equation}
	T^{OV}_{dom}(K_1,K_2,K_3,K_4) =
                \frac{3}{8\pi^3} \int_{0}^{\eta_{0}}d\eta
                \left( \frac{ aD\dot{D}}{ D_0^2 } \right)^4 
                \frac{g^4(\eta)}{\eta^6} \sum_{i,j,l=1;i \neq j \neq l}^{n=4}
                f \left(
                     \frac{\vec{K}_i}{\eta},
                     \frac{\vec{K}_j}{\eta},
                     \frac{\vec{K}_l}{\eta}
                  \right)
\label{eq:trispectrum_dom2}
\end{equation}
where $f$ is given by (\ref{eq:f_tris7}). The power of the Limber
approximation was to reduce an almost impossible integration to a 
$4$-dimensional integral, which can be numerically calculated 
for a chosen configuration of the wavenumbers $K$. 
\subsubsection{Nonlinear extension: the kSZ effect trispectrum}
We follow the same approach as for the power spectrum and bispectrum and 
again we calculate the nonlinear extension for the dominant contribution.
As we have four CDM power spectrum showing up
in the expression for the trispectrum, which we need to divide among 
linear/nonlinear contributions from density/velocity contributions we will  
assume the following trispectrum statistic
\begin{equation}
	T^{OV-NL}_{dom}(K_1,K_2,K_3,K_4) =
                \frac{3}{8\pi^3} \int_{0}^{\eta_{0}}d\eta
                \left( \frac{ aD\dot{D}}{ D_0^2 } \right)^4 
                \frac{g^4(\eta)}{\eta^6} 
	        \sum_{i,j,l,k=1;i \neq j \neq l}^{n=4}
                \left( \frac{P^{NL}(K_k/\eta)}{P(K_k/\eta)} \right)^{2}
                f \left(
                     \frac{\vec{K}_i}{\eta},
                     \frac{\vec{K}_j}{\eta},
                     \frac{\vec{K}_l}{\eta}
                  \right) .
\label{eq:trispectrumNL_dom}
\end{equation}
\subsubsection{Signal-to-noise} \label{subsec:S_N_trispectrum}
As pointed out by Zaldarriaga~\cite{Zaldarriaga00} and later by 
Hu~\cite{Hu99}, the maximal signal-to-noise can be proven to be
\begin{equation}
\chi^2 \equiv  \left( \frac{S}{N} \right)^2
       = f_{sky} \sum_{L}\sum_{\ell_1 \leq \ell_2 \leq \ell_3 \leq \ell_4} \frac{1}{2L+1}
         \frac{\mid T_{\ell_1 \ell_2 \ell_3 \ell_4} (L)\mid^2}
              {C_{\ell_1}^{tot}C_{\ell_2}^{tot}C_{\ell_3}^{tot}C_{\ell_4}^{tot} }  
\label{eq:chi_tri}
\end{equation}
where $T_{\ell_1 \ell_2 \ell_3 \ell_4} (L)$ is one of the possible 
configurations of the full-sky trispectrum as defined 
in what follows. The covariance matrix used to obtain the Fisher matrix is 
calculated assuming full-sky coverage and $\ell_1 \leq \ell_2 < \ell_3 \leq \ell_4$
by Komatsu~\cite{Komatsu02} in the weakly non-Gaussian limit. If 
$\ell_1 \leq \ell_2 < \ell_3 \leq \ell_4$ is not respected, the covariance 
would be distributed across many $L$s and can lead to 
overestimates of the signal-to-noise~\cite{Hu01}. By not respecting
this last constrain in Eq.~(\ref{eq:chi_tri}) we are 
calculating an upper limit of the $S/N$ estimate.

Concerning the equivalence between the full-sky and flat-sky formalisms,
we follow Hu's Appendix~\cite{Hu01} where it is argued that we can find 
a relation between the two formalisms by breaking up the trispectrum in
the three possible combinations in each configuration defined by 
$(\ell_1,\ell_2,\ell_3,\ell_4)$ 
\begin{equation}
    T_{\ell_1 \ell_2 \ell_3 \ell_4}    = T_{(\ell_1 \ell_2) (\ell_3 \ell_4)}(L_{12}) 
                            +T_{(\ell_1 \ell_3) (\ell_2 \ell_4)}(L_{13}) 
                            +T_{(\ell_1 \ell_4) (\ell_3 \ell_2)}(L_{14}) 	
\label{eq:trisp_dec}
\end{equation}
where $L_{ij}$ correspond to the side of the triangle of sides $(\ell_i, \ell_j, L_{ij})$.
Each of the $T_{(\ell_i \ell_j) (\ell_k \ell_m)}(L_{ij})$ is then related to the 
flat-sky equivalent by
\begin{eqnarray}
    T_{(\ell_i \ell_j) (\ell_k \ell_m)}(L_{ij}) 
        &=& \frac{2L+1}{4\pi}\sqrt{(2\ell_i+1)(2\ell_j+1)(2\ell_l+1)(2\ell_m+1)}
            \left( 
            \begin{array}{ccc} \ell_i & \ell_j & L_{ij} \\ 
                                0  &  0  &  0 
            \end{array} 
            \right)
            \left( 
            \begin{array}{ccc} \ell_k & \ell_m & L_{ij} \\ 
                                0  &  0  &  0 
            \end{array} 
            \right)\nonumber \\
        & & T( (K_i=\ell_i,K_j=\ell_j),(K_k=\ell_k,K_m=\ell_m) (L_{ij}) ) .
\label{eq:tris_full_flat}
\end{eqnarray}
As for the bispectrum, the Wigner-3j vanishes if $\ell_i+\ell_j+L_{ij}=odd$. 
So the full-sky trispectrum can only be estimated for even terms.
\subsubsection{Results} \label{subsec:results_trispectrum}
We are interested in the numerical evaluation of the dominant 
contribution of the trispectrum as it may be of cosmological 
interest in the near future due to the next generation of experiments.
We choose the trapezoidal configuration
in Fourier space $K_1=K_2=K_3=K_4=l$ with an angle 
$\theta$ between two consecutive sides. 
This will give rise to the following flat-sky trispectrum
\begin{eqnarray}
	T^{OV}_{dom}(\ell,\theta) 
             &=&\frac{3}{2\pi^3} \int_{0}^{\eta_{0}}d\eta
                \left( \frac{ aD\dot{D}}{ D_0^2 } \right)^4 
                \frac{g^4(\eta)}{\eta^6}\nonumber\\ 
             & & ( f \left(\frac{\ell}{\eta},-1,\epsilon,-\epsilon)\right)
                  +f \left(\frac{\ell}{\eta},-1,-\epsilon,\epsilon)\right)
                  +f \left(\frac{\ell}{\eta},\epsilon,-1,-\epsilon)\right)\nonumber\\ 
             & &  +f \left(\frac{\ell}{\eta},-\epsilon,-1,\epsilon)\right)
                  +f \left(\frac{\ell}{\eta},\epsilon,-\epsilon,-1)\right)
                  +f \left(\frac{\ell}{\eta},-\epsilon,\epsilon,-1)\right)
               )
\label{eq:trispectrum_dom_red}
\end{eqnarray}
where $\epsilon=\cos(\theta)$ and $f$ is defined by Eq.~(\ref{eq:f_tris7}).
The various $f$ correspond to the different specific configurations due to the
ordering of the vectors $\vec{K}$s in the sum over the function $f$ in
Eq.~(\ref{eq:trispectrum_dom2}).
Note that the analytical expression of $T^{OV}_{dom}$ is symmetric 
around $\theta = \pi/2$ reflecting the symmetry of the configuration.
The nonlinear analog follows from the previous equation. 

With this configuration which depends only on the multipole $\ell$ and 
the angle $\theta$ between two sides 
one can calculate an 
estimate of the order of magnitude of the $(S/N)^2(\theta)$ per bin of $\ell$ 
were we to include all the remaining $\ell$ modes
of the full-sky trispectrum. 
Indeed in the Eq.~(\ref{eq:chi_tri}), 
we see that the number of modes contributing to the $(S/N)^2$ per bin of $\ell$ increases
as $\ell^{3}$ so we will have the following relation for $\chi^2$ 
in a bin of l
\begin{equation}
\frac{d\chi^2(\theta)}{d\ell}  \sim \ell^{3} f_{sky}\,\sum_{i=1}^{n=3}
                                       \frac{\mid T_{\ell} (L_i)\mid^2}{(2L_i+1)(C_{\ell}^{tot})^4}
\label{eq:chi_trisp_est}
\end{equation}
where $T_{\ell} (L_i)$ is one of the $3$ possible configurations
of the full-sky trispectrum for a given multipole $\ell$ and angle $\theta$
which needs to be calculated
from its flat-sky counterpart (Eq.~(\ref{eq:trispectrum_dom_red}))
using Eqs.~(\ref{eq:trisp_dec}) and~(\ref{eq:tris_full_flat}).
The $C_{\ell}^{tot}$ is calculated as for the bispectrum (see 
Eq.~(\ref{eq:cl_tot})). Again, $f_{sky}$ 
represents the reduction in signal-to-noise due to incomplete sky coverage. 
\vspace{4mm}

\begin{figure*}
\centering 
\vbox to 0.30\textheight{
{\hbox to \textwidth{
\null\null
\epsfxsize=0.45\textwidth
\epsfysize=0.30\textheight
\epsfbox{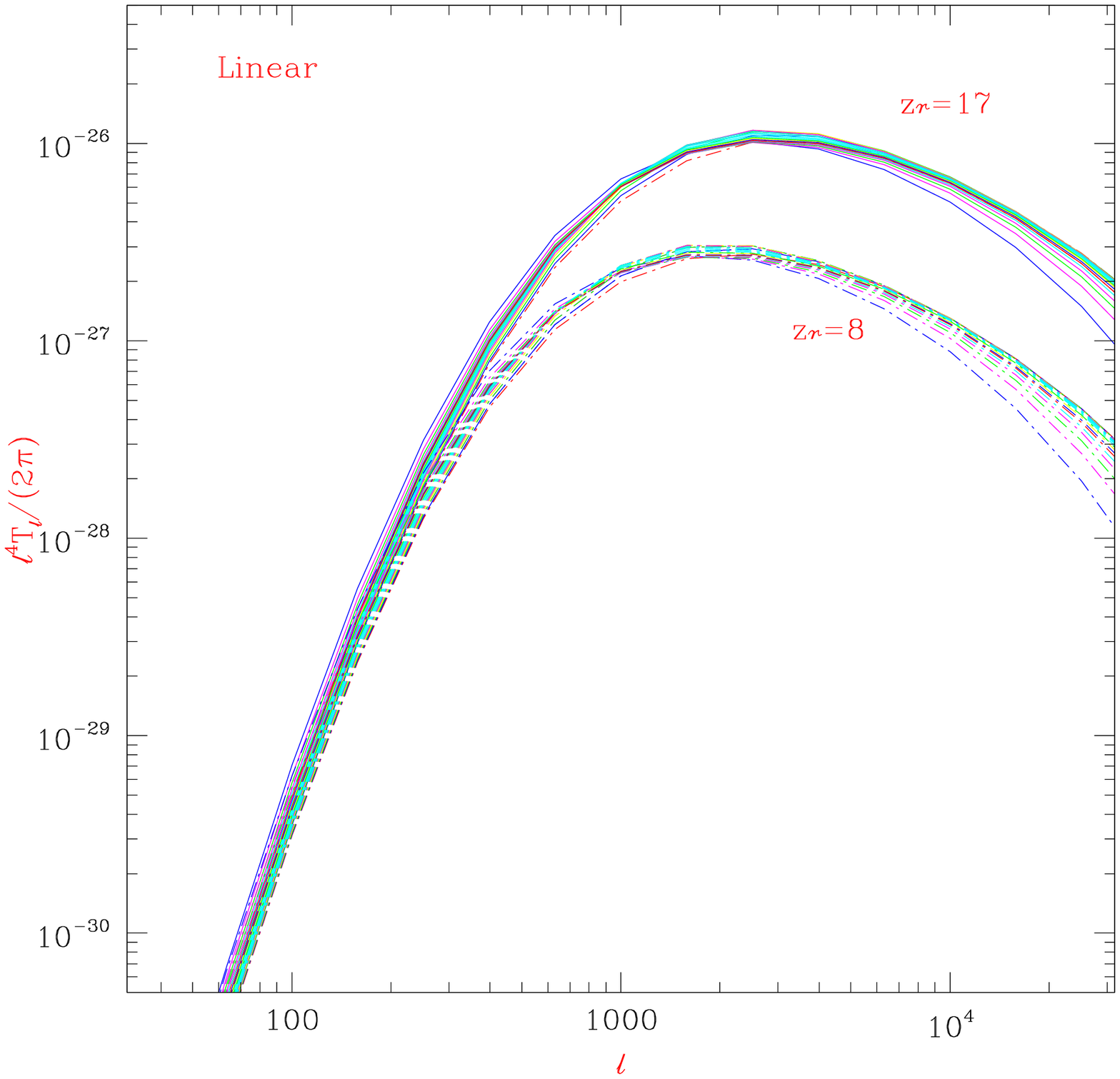}
\hfill
\epsfxsize=0.45\textwidth
\epsfysize=0.30\textheight
\epsfbox{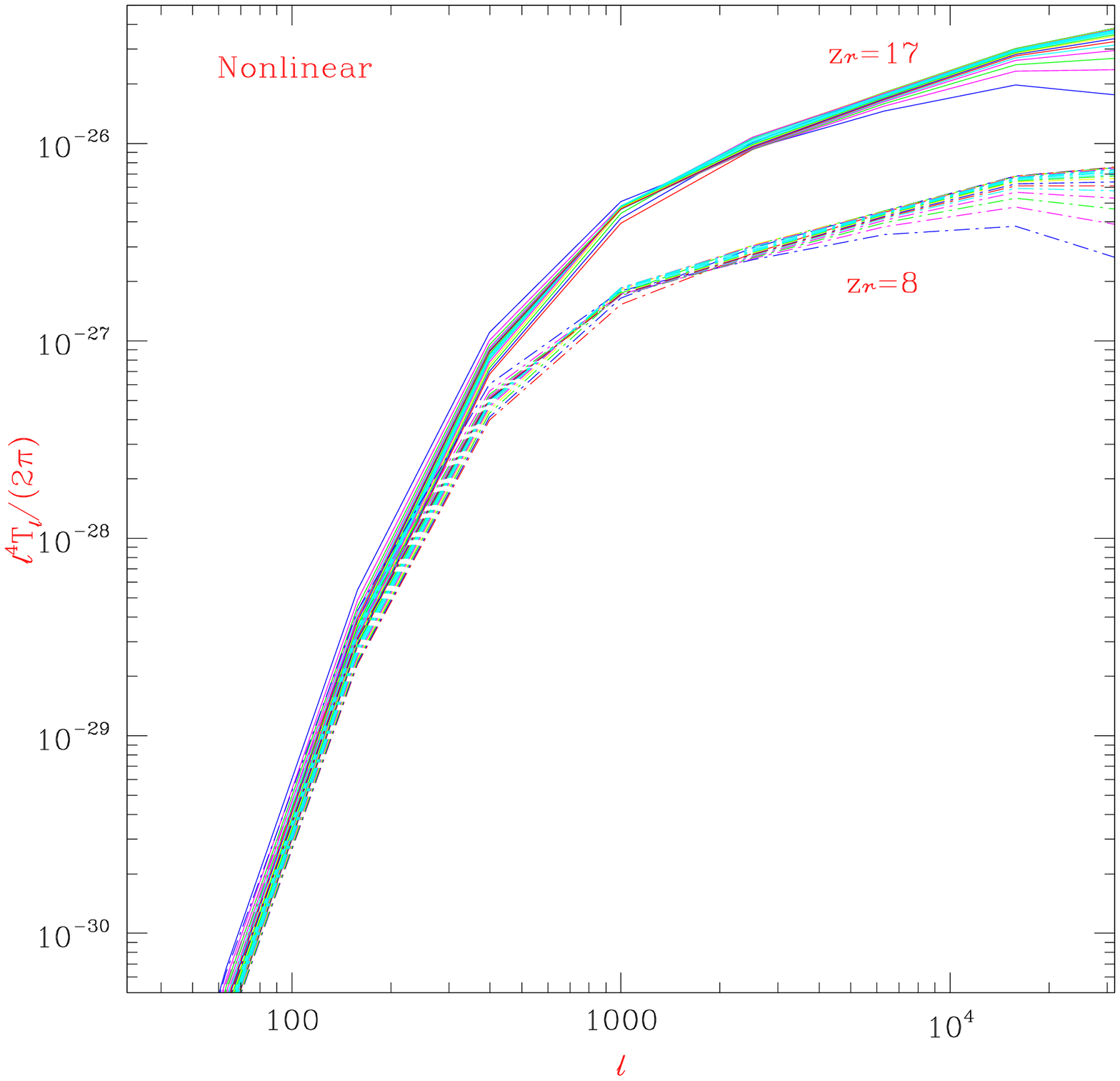}
\hfill
}}}
\caption{ \small{ 
{\it Left panel---} Linear 
and {\it Right panel ---} nonlinear 
flat-sky trispectrum of the OV effect for
geometrical configurations such that 
$ -0.95 \leq \epsilon \leq 0.00$ in steps of $0.05$.
The amplitude of the trispectrum decreases as 
$\epsilon$ decreases from $0.00$ to $-0.95$. Because
the power is symmetric in $\epsilon$ around $0.00$ we 
only plotted the negative $\epsilon$s. 
All the plots were calculated for the fiducial $\Lambda$CDM model. The 
{\it dot dash} lines correspond to $z_r=8$ and 
the {\it solid} lines to $z_r=17$. We assumed
$\Delta z_r= 0.1(1+z_r)$.
}}
\label{fig:trispec_l_nl}
\null
\end{figure*}

We show the plots for the linear and the nonlinear OV flat-sky 
trispectrum for $-0.95 < \epsilon < 0.00$, assuming both $z_r=8$
and $z_r=17$, in figure~(\ref{fig:trispec_l_nl}).
As the trispectrum is symmetric around $\epsilon = 0.00$ we only plotted
the negative $\epsilon$s, but this choice was arbitrary.
For $z_r=8$ the linear trispectrum peaks around multipole $\ell \simeq 2.10^3$
and has a maximum amplitude of $\ell^4T(\ell)/(2\pi) \simeq 2.4\times10^{-27}$,
regardless of the configuration. Choosing an earlier reionization
of $z_r=17$ shifts the peak to $\ell \simeq 3.10^3$ 
and increases the overall amplitude of the trispectrum to 
$l^4T(\ell)/(2\pi) \simeq 1.2\times10^{-26}$. Numerically we find for the amplitude 
of the trispectrum the approximate scaling dependence 
with reionization history 
$T(\ell \simeq \,170) \simeq 2.6\times10^{-31}\,x_e^4\,\log^{0.8}{(1+z_r)}$,
in agreement with the relation found for the power spectrum.
The effect of increasing
the value of angle between two consecutive sides of the trapezoid
is to increase the power at small scales ($\ell > 10^4$), 
such that only small scales 
are sensitive enough to probe different configurations. 
At those small scales, the power is maximum for $\epsilon = 0.00$
and then decreases as we decrease the angle. So at small angular 
scales the square configuration is the one contributing the most. 
We could have expected this behavior as the OV effect has
a quite symmetric morphological signature, such that 
a more {\it filamentary} structure probed by the collapsed 
configuration of the trispectrum is not as likely.  

Here we do not observe the sharp drop in power at small scales
due to the Limber cancellation which enables one to speak of
the trispectrum of the kSZ effect.
Indeed, concerning the nonlinear trispectrum, we observe an
interesting feature. Contrary to the bispectrum and
similarly to the power spectrum, the trispectrum is strongly
affected by the weak nonlinear enhancement due to formation
of structure at small scales. This enhancement has 
the power to broaden the shape, to shift the peak of the effect 
to smaller angular scales and 
to increase slightly the amplitude in a configuration dependent way.
So (for $z_r=8$/$z_r=17$) we measure a higher and higher maximum amplitude
($\simeq 2.4-7.2\times10^{-27}$/$\simeq 1.9-4.8\times10^{-26}$) 
at a smaller and smaller angle ($\ell \simeq 1-4.10^4$)
when you go from the collapsed trapezoid to the square 
configuration. It is worth noticing that, whereas the amplitude
of the nonlinear trispectra increases, the multipoles corresponding to 
the peak of the effect remain the same
for both reionization scenarios. This is because 
the nonlinearities take place at the same instant in time 
for both reionization scenarios, leaving an imprint at the same
characteristic angular scale.
 
\begin{figure}[t]
\centerline{\epsfxsize=3.5truein\epsffile{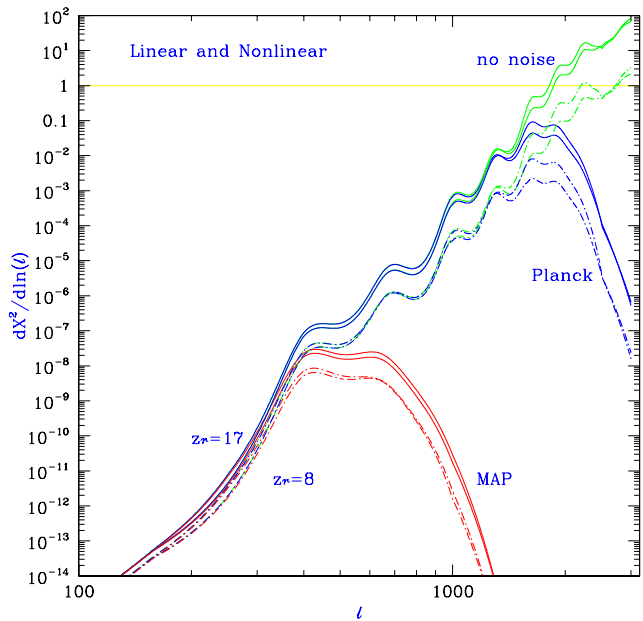}}
\caption{ \small{
Contribution to the $\chi^2$ per log interval 
in $\ell$ for $\epsilon = -0.95$ for the OV full-sky linear/nonlinear trispectrum 
with no instrumental noise ({\it top}), 
Planck noise ({\it middle}) and MAP noise ({\it bottom}) included 
in the variance. Again we used the specifications in the table~\ref{table1}.
All the plots were calculated for the fiducial $\Lambda$CDM model. The 
{\it dot dash} lines correspond to $z_r=8$ and 
the {\it solid} lines to $z_r=17$. We assumed
$\Delta z_r= 0.1(1+z_r)$.
The higher amplitudes for each of the experiments correspond 
to the contributions from the full-sky nonlinear trispectrum. 
The horizontal line at $d\chi^2/dln(l)=1$
shows the minimum detection threshold. 
The total $S/N$ for the OV full-sky linear trispectrum for MAP, Planck 
and a perfect experiment respectively are: 
$1.4\times10^{-3}$, $1.1$ and $27.4$ assuming $z_r=8$,
and $2.9\times10^{-3}$, $4.7$ and $149$ assuming $z_r=17$.
}}
\label{fig:chi_tri}
\end{figure}

But the most important quantity is the signal-to-noise. 
We show in figure~(\ref{fig:chi_tri}) the estimated contributions
to $\chi^2$ per log interval in $\ell$
for the linear and the nonlinear OV full-sky 
trispectrum ($z_r=8$ and $z_r=17$) had we included 
all the modes for $\epsilon=-0.95$
with no instrumental noise, Planck noise and MAP noise included 
in the variance~\cite{MAP/PLANCK}.  
We chose $\epsilon=-0.95$, i.e. the collapsed configuration,
because it generates the highest contributions 
to the full-sky trispectrum near the peak of the effect, 
contrary to the flat-sky trispectrum
for which the square configuration was the one producing the
highest amplitude. The square configuration continues to
provide the strongest contribution at very small angular scales,
but at larger angular scales the collapsed configuration 
dominates. This is due to the angular averaged factors 
relating flat- to full-sky trispectra 
(see Eq.~(\ref{eq:tris_full_flat})).
Though it is not explicit in the figure, the 
$d\chi^2/d\ln{\ell}$ is zero when $2\ell+L$ is odd.
Again, as for the bispectrum case, the structure in the $d\chi^2/d\ell$
arises mainly from the structure of the CMB primary
power spectrum at $\ell > 200$. The continuous rise 
at $\ell\simeq 3000$ for a perfect experiment is due to the fact that
the signal decreases slower than the contributions to the $C_{\ell}$
from primary, OV and thermal SZ anisotropies up to these scales.
Contrary to the common and most naive expectation, 
for Planck, an eventual detection is possible 
at arcminute scales. Indeed, 
the total $S/N \sim 4.7$ for $z_r=17$. 
Most of the contributions come
from multipoles between $\ell=1-2.10^3$. This probably 
is the most important conclusion of
this work and illustrates our predictions.
This result should be taken with caution as it corresponds to
a very optimistic upper limit on the $S/N$ 
(see sections Sec.~\ref{subsec:Signal-to-noise} and 
Sec.~\ref{subsec:S_N_trispectrum}). Firstly, the use of the 
Fisher matrix formalism gives the minimum variance for our 
statistic (see section Sec.~\ref{subsec:Signal-to-noise}). 
Secondly, the $\chi^2$ method assumes that the form 
of the model is correct, which may not be the case. Thirdly,
when calculating the covariance matrix of the Fisher matrix,
two simplifying assumptions were used:
that the main contribution to the covariance was Gaussian in nature
and that we observed with full-sky coverage. 
Finally, other physical mechanisms, like for example lensing effects or
the thermal SZ effect~\cite{Hu01,Cooray01b,Zaldarriaga00,Cooray:Kesden02}, 
and unaccounted foregrounds~\cite{Tegmark99} 
can contribute to the trispectrum at this level 
and so the separability problem needs to be addressed 
in due time. Of course, uncertainties of roughly an order of magnitude 
in the modeling of the
thermal SZ signal are also a source of error in our estimates. 
The forecasted ability of future multi-frequency experiments 
to remove most of the thermal SZ contributions would minimize 
these uncertainties and would much favor a detection. 
Last but not least, further progress 
in the implementation of optimal unbiased trispectrum estimators
to probe such small scales and power is required. 
\section{Conclusions} \label{sec:Conclusions}
Due to its strong predictive power, linear theory 
is a very sensitive probe of the 
early stages of the reionization history through 
the Ostriker and Vishniac effect. Analytical expressions 
for its correlation functions can be derived and
their measurement would be of high value to 
our present knowledge of that still unclear 
epoch of the universe evolution.
We have presented detailed calculations of the three Fourier 
statistics of interest of the OV effect, 
the power spectrum, the bispectrum and the trispectrum. For
that purpose we have developed a new technique that allows 
one to obtain their dominant contributions
under the Limber approximation framework. This method
is applicable to the derivation of any statistics involving 
correlations among vector-like effects. It illustrates
what was expected under statistical homogeneity and isotropy
assumptions and the vector and small scale nature of the OV effect.
We also evaluated numerically, as a function of scale and
for a specific configuration (equilateral for the bispectrum 
and trapezoidal for the trispectrum), these statistics for a flat
$\Lambda$CDM cosmology and two reionization scenarios. The first one is
based on our pre-WMAP knowledge ($z_r=8$) and the second one takes into account
the high values for the electron optical depth measured by WMAP ($z_r=17$).
We numerically obtained approximate scaling relations for the amplitude of 
the OV statistics on the reionization history considered and found
that the dependence is stronger on the ionization fraction than on the
redshift of reionization. 
We have also studied their detectability in view of future 
satellite experiments. The
alternation of dominant/subdominant/dominant higher 
order correlation functions was numerically shown. 
While the bispectrum is undetectable even by a 
perfect multi-frequency experiment capable of subtracting
the thermal SZ contributions,
the trispectrum could
be measured by Planck or by interferometer 
experiments targeting
arcminute scales with high sensitivity and for
a sufficiently long period of time.  
This provides a unique signal distinguishing 
the OV effect from other non vector-like 
secondary anisotropies and could be useful when trying 
to separate different physical mechanisms imprinting
themselves on these measurable statistics. 
One should bear in mind that despite this useful 
characteristic signature, our results are quite
optimistic, although encouraging, as they rely on various
analytically helpful idealized assumptions, as described previously. 
Also, other contributions to the signal are to be 
expected at the arcminute level and thus further study  
of CMB small-scale secondary anisotropies
and foregrounds contributions to 
the trispectrum is required and much justified.
In order to obtain an upper limit on the possible kSZ
contributions, we also extended 
our calculations to the mildly nonlinear 
regime. We found that, contrary to the bispectrum, 
there is a noticeable enhancement 
of the contributions of the trispectrum in a 
morphological dependent way and that this enhancement
reflects itself on the calculations of 
the signal-to-noise. Hence nonlinearities are expected
to enhance the even non-Gaussian signals produced by the
OV effect and further complicate its disentanglement from  
inevitable model-dependent nonlinear effects arising from 
structure formation. Conversely, having a template of the OV 
effect can help in extracting the nonlinear contributions
to reionization at those angular scales, providing another
possible window on the complex physics associated with 
reionization.
\begin{acknowledgements}
I am much indebted to Pedro G. Ferreira 
whose suggestions, comments and constant support have
been invaluable during the preparation of this work. 
I thank also Asantha Cooray for very helpful discussions and
comments and for the thermal SZ data used to facilitate direct comparison
with his results, Evan Scannapieco for helping 
to improve the manuscript and to correct
a fundamental error in the calculations, 
Nabila Aghanim for her support and the thermal SZ data 
used throughout the article, Andrew Jaffe for his comments 
and Martin Kunz for his 
encouragement. This work was supported by 
the Funda\c{c}\~{a}o para a Ci\^{e}ncia e a Tecnologia under
the reference PRAXIS XXI/BD/21249/99.
\end{acknowledgements}
\appendix 
\section{Limber Approximation} \label{sec:appendixut}
We review the fundamental steps of the limber approximation as used
in the text. The Limber Eq.~\cite{Limber54} describes the 
two-point statistics of a field which is the two-dimensional projection 
on the sky of a three-dimensional field whose statistical properties vary 
slowly along the line of sight. The Fourier space analog of this result 
was calculated by Kaiser~\cite{Kaiser92}. Later it was further extended
from the flat-sky approximation to an all-sky approach for spatially flat 
cosmologies by Hu and White~\cite{Hu:White96} and to open cosmologies by Hu
\cite{Hu99}. Buchalter \emph{et. al.}~\cite{Buchalter00} derived the Fourier space 
analog of the Limber's equation for the bispectrum.

Here we review the bispectrum derivation of Buchalter \emph{et. al.} and generalize 
it to the trispectrum. This applies as well to higher-order statistics. 
The error introduced by the Limber approximation is inferior to $1\%$ 
for effects with $l>200$~\cite{Cooray:Hu00,Pollo01}. 

Suppose we observe the projection along the line of sight of a three-dimensional 
statistically homogeneous and isotropic random field $f$
\begin{equation}
p(\vec{\theta})=\int_{0}^{\infty}d\eta q(\eta)f(\eta \hat{\theta})
\label{eq:Limber_1}
\end{equation}
where $\vec{\theta}=(\theta_1,\theta_2,0)$ and $\hat{\theta}=(\theta_1,\theta_2,1)$.
We propose ourselves to find a relation between the spatial bispectrum 
$B_p(K_1,K_2,K_3)$ of $p$ and the spatial bispectrum $B_f(K_1,K_2,K_3)$ of $f$.
Following Kaiser method, we consider $p$ to be the sum of the contributions 
from narrow shells with a width $\Delta \eta$ much bigger than the relevant 
wavelength, that is $\theta<<\Delta \eta / \eta<<1$. This choice allows us 
to look at fluctuations on scales much less than the characteristic scale 
over which $q$ varies and to assume that contributions from different shells 
are statistically independent. We then calculate the contributions to the 
bispectrum from each of the shells. At the end we can sum the power for all 
the shells relying on their statistical independence.   

Assuming that $q$ varies very little along the shell and that the section of 
the shell is plane-symmetric the contribution from the shell of width 
$\Delta \eta$ centered at $\eta_0$ is
\begin{equation}
\Delta p(\vec{\theta})=
     q(\eta_0)\int_{\eta_0-\Delta \eta/2}^{\eta_0+\Delta \eta/2}d\eta 
                      f(\eta_0 \theta_1,\eta_0 \theta_2, \eta)
\label{eq:Limber_2}
\end{equation}
Decomposing the fields in Fourier space, the spectrum of $\Delta p$ is
\begin{equation}
\tilde{\Delta p}(\vec{K})=q(\eta_0)\int \frac{d^3k}{(2\pi)^3} 
       \tilde{f}({\mathbf k})\int d^2\theta e^{i(\eta_0 \vec{k}-\vec{K}).\vec{\theta}} 
        \int_{\eta_0-\Delta \eta/2}^{\eta_0+\Delta \eta/2}d\eta e^{ik_z\eta}
\label{eq:Limber_3}
\end{equation}
where $\vec{K}=(K_x,K_y,0)$ and ${\mathbf k}=(k_x,k_y,k_z)=(\vec{k},k_z)$.
We can perform the $d^2\theta$ integral which is 
$(2\pi)^2 / \eta_0^2 \delta_D^2 (\vec{k}-\vec{K} / \eta_0)$ and the time integral 
which is $\Delta \eta j_0(k_z\Delta \eta / 2)$ to yield
\begin{equation}
\tilde{\Delta p}(\vec{K})=\frac{\Delta \eta q(\eta_0)}{\eta_0^2} \int \frac{dk_z}{(2\pi)}\tilde{f}(\frac{K_x}{\eta_0},\frac{K_y}{\eta_0},k_z)j_0(k_z\Delta \eta/2)
\label{eq:Limber_4}
\end{equation}
Using 
$<\tilde{f}({\mathbf k}_1)\tilde{f}({\mathbf k}_2)\tilde{f}({\mathbf k}_3)>=(2\pi)^3B_f(k_1,k_2,k_3)\delta_D^3({\mathbf k}_1+{\mathbf k}_2+{\mathbf k}_3)$ the three-point spectrum is
\begin{eqnarray}
<\tilde{\Delta p}(\vec{K}_1)\tilde{\Delta p}(\vec{K}_2)\tilde{\Delta p}(\vec{K}_3)>
    &=& \frac{\Delta\eta^3q^3(\eta_0)}{\eta_0^6}
	\int dk_{1z}
	\int dk_{2z}
	\int dk_{3z}
	B_f(\sqrt{\frac{K_1^2}{\eta_0^2}+k_{1z}^2},
    	\sqrt{\frac{K_2^2}{\eta_0^2}+k_{2z}^2},
    	\sqrt{\frac{K_3^2}{\eta_0^2}+k_{3z}^2})\nonumber \\
    & &	j_0(k_{1z}\Delta\eta/2)j_0(k_{2z}\Delta\eta/2)j_0(k_{3z}\Delta\eta/2)
    	\delta_D^2(\vec{K}_1/\eta_0+\vec{K}_2/\eta_0+\vec{K}_3/\eta_0)
     	\delta_D(k_{1z}+k_{2z}+k_{3z})\nonumber\\
    &=& \frac{\Delta \eta^3 q^3(\eta_0)}{\eta_0^4}
	\int dk_{1z}
	\int dk_{2z}
	B_f(\sqrt{\frac{K_1^2}{\eta_0^2}+k_{1z}^2},
    	\sqrt{\frac{K_2^2}{\eta_0^2}+k_{2z}^2},
    	\sqrt{\frac{K_3^2}{\eta_0^2}+(k_{1z}+k_{2z})^2})\nonumber\\
    & &	j_0(k_{1z}\Delta\eta/2)j_0(k_{2z}\Delta\eta/2)j_0((-k_{1z}-k_{2z})\Delta\eta/2)
    	\delta_D^2(\vec{K}_1+\vec{K}_2+\vec{K}_3)
\label{eq:Limber_5}
\end{eqnarray}
The important simplification comes from the fact that the major contribution from 
the first two Bessel functions comes from $k_{1z} < 1/\Delta \eta$ and 
$k_{2z} < 1/\Delta \eta$. But by assumption $K_1/\eta_0>>1/\Delta \eta$, 
$K_2/\eta_0>>1/\Delta \eta$ and $K_3/\eta_0>>1/\Delta \eta$. Therefore, 
$k_{1z} << K_1/\eta_0$, $k_{2z} << K_2/\eta_0$ and $k_{1z}+k_{2z} << K_3/\eta_0$. 
So we can neglect all Fourier modes parallel to the line of sight. To a very 
good approximation $B_f(\sqrt{\frac{K_1^2}{\eta_0^2}+k_{1z}^2},\sqrt{\frac{K_2^2}{\eta_0^2}+k_{2z}^2},\sqrt{\frac{K_3^2}{\eta_0^2}+(k_{1z}+k_{2z})^2}) \simeq B_f(\frac{K_1}{\eta_0},\frac{K_2}{\eta_0},\frac{K_3}{\eta_0})$.
The integration of the Bessel functions gives
\begin{equation} 
\int du \int dv j_0(u)j_0(v)j_0(u+v)=\pi^2
\label{eq:Limber_6}
\end{equation}
We obtain finally
\begin{eqnarray}
<\tilde{\Delta p}(\vec{K}_1)\tilde{\Delta p}(\vec{K}_2)\tilde{\Delta p}(\vec{K}_3)>
    &=& 4 \pi^2 \frac{\Delta\eta q^3(\eta_0)}{\eta_0^4}
	B_f(\frac{K_1}{\eta_0},\frac{K_2}{\eta_0},\frac{K_3}{\eta_0})
    	\delta_D^2(\vec{K}_1+\vec{K}_2+\vec{K}_3)
\label{eq:Limber_7}
\end{eqnarray}
Summing over the shells, and using $<\tilde{p}(\vec{K}_1)\tilde{p}(\vec{K}_2)\tilde{p}(\vec{K}_3)>=(2\pi)^2B_p(K_1,K_2,K_3)\delta_D^2(\vec{K}_1+\vec{K}_2+\vec{K}_3) $ 
\begin{eqnarray}
B_p(K_1,K_2,K_3)
    &=& \int d\eta \frac{q^3(\eta)}{\eta^4}
	B_f(\frac{K_1}{\eta},\frac{K_2}{\eta},\frac{K_3}{\eta})
\label{eq:Limber_8}
\end{eqnarray}
The exact same reasoning can be applied to the calculation of the 
trispectrum. This time 
we obtain for the same projection
\begin{eqnarray}
T_p(K_1,K_2,K_3,K_4)    
&=&    \int d\eta \frac{q^4(\eta)}{\eta^6}
	T_f(\frac{K_1}{\eta},\frac{K_2}{\eta},\frac{K_3}{\eta},\frac{K_4}{\eta})
\label{eq:Limber_9}
\end{eqnarray}


\begin{thebibliography}{99}
\bibitem{MAP/PLANCK} See the MAP homepage at {\tt http://map.gsfc.nasa.gov},
         and the Planck homepage at {\tt
         http://astro.estec.esa.nl/SA-general/Projects/Cobras/cobras.html}.
\bibitem{MINT} See the MINT homepage {\tt http://imogen.princeton.edu/mintweb/}
\bibitem{CBI} See the CBI homepage {\tt http://www.astro.caltech.edu/~tjp/CBI/}
\bibitem{ACT} See the ACT homepage {\tt http://www.hep.upenn.edu/~angelica/act/science.html}
\bibitem{Hu01} W. Hu, 2001, Phys. Rev. D, 64, 083005
\bibitem{Komatsu:Spergel00} E. Komatsu and D. Spergel, 2001, Phys. Rev. D, 63, 063002
\bibitem{Kunz01} M. Kunz, A. J. Banday, P. G. Castro, P. G. Ferreira 
         and K. M. Gorski, 2001, Astrophys. J. Lett., 563, 99
\bibitem{Santos01} M. Santos \emph{et. al.}, 2002, Phys. Rev. Lett., 88, 241302
\bibitem{Komatsu01} E. Komatsu \emph{et. al.}, 2002, Astrophys. J., 566, 19
\bibitem{Troia03} G. De Troia \emph{et. al.}, 2003, astro-ph/0301294
\bibitem{Ost:Vish86} J. P. Ostriker and E. T. Vishniac, 1986, Nature, 322, 804
\bibitem{Hu94} W. Hu, D. Scott and J. Silk, 1994,  Phys. Rev. D, 49, 648
\bibitem{Dod:Jubas93} S. Dodelson and J. M. Jubas, 1995, Astrophys. J., 439, 503
\bibitem{Vishniac87} E. T. Vishniac, 1987, Astrophys. J., 322, 597
\bibitem{Peebles} See, e.g., P. J. E. Peebles, {\sl Principles
         of Physical Cosmology} (Princeton University Press,
         Princeton, 1993).
\bibitem{Hu:Sugiyama96}  W. Hu and N. Sugiyama, 1996, Astrophys. J., 471, 542
\bibitem{Bardeen86} J. M. Bardeen \emph{et. al.}, 1986, Astrophys. J., 304, 15
\bibitem{Eisenstein:Hu99} D. J. Eisenstein and W. Hu, 1999, Astrophys. J., 511, 5
\bibitem{Bunn:White97} E. F. Bunn and M. White, 1997, Astrophys. J., 480, 6
\bibitem{Barkana:Loeb01} A. Loeb and R. Barkana, 2001, Annu. Rev. Astron. Astrophys., 
         39, 19L
\bibitem{Haiman:Knox99} Z. Haiman and L. Knox, 1999, \emph{Microwave Foregrounds}, 
         ASP Conference Series 181, eds. A. de Oliveira-Costa and M. Tegmark, p.227
\bibitem{Bernardis00} P. de Bernardis \emph{et. al.}, 2002, Astrophys. J., 564, 559
\bibitem{Hanany00} S. Hanany \emph{et. al.}, 2001, Am. Astron. Soc., 199.3403
\bibitem{Netterfield01} C. B. Netterfield \emph{et. al.}, 2002, Astrophys. J., 
         571, 604
\bibitem{Griffiths99} L. Griffiths  \emph{et. al.}, 1999, Mon. Not. R. Astron. Soc., 
         308, 854
\bibitem{Tegmark00} M. Tegmark and M. Zaldarriaga, 2000, Phys. Rev. Lett., 85, 2240
\bibitem{Hu:Holder03} W. Hu and G. P. Holder, 2003, astro-ph/0303400
\bibitem{Kogutetal03} A. Kogut \emph{et. al.}, 2003, astro-ph/0302213
\bibitem{Spergeletal03} D. Spergel \emph{et. al.}, 2003, astro-ph/0302209
\bibitem{Songaila:Cowie02} A. Songoila, L. L. Cowie, 2002, Astron. J., 123, 2183
\bibitem{Djorgovski01}  S. G. Djorgovski, S. M. Castro, D. Stern and A. Mahabal, 
         2001, Astrophys. J. Lett., 560, 5D
\bibitem{Gunn:Peterson65} J. E. Gunn and B. A. Peterson, 1965, Astrophys. J.,
         142, 1633
\bibitem{Becker01} R. H. Becker \emph{et. al.}, 2001, Astron. J., 122, 2850
\bibitem{Gnedin01} N. Y. Gnedin, 2001, Submitted to Mon. Not. R. Astron. Soc., 
         astro-ph/0110290
\bibitem{Haiman&Holder03} Z. Haiman and G. P. Holder, 2003, astro-ph/0302403
\bibitem{Hui&Haiman03} L. Hui and Z. Haiman, 2003, astro-ph/0302439
\bibitem{Wyithe&Loeb03} S. Wyithe and A. Loeb, 2003, astro-ph/0302297
\bibitem{Sunyaev78} R. A. Sunyaev , 1978, 
         \emph{Large-Scale Structure of the Universe}, Dordrecht, D. Reidel
         Publishing Co., eds. M. S. Longair and J. Einasto, p.409
\bibitem{Kaiser84} N. Kaiser, 1984, Astrophys. J., 282, 374
\bibitem{Bruscoli99} M. Bruscoli \emph{et. al.}, 2000, Mon. Not. R. Astron. Soc., 
         318, 1068
\bibitem{Ma:Fry01} Chung-Pei Ma and J. N. Fry, 2002, Phys. Rev. Lett., 88, 211301
\bibitem{Refregier:Teyssier00} A. Refregier and R. Teyssier, 2000, 
         Submitted to Phys. Rev. D, astro-ph/0012086
\bibitem{Scaramella93} R. Scaramella, R. Cen and J. P. Ostriker, 1993, 
         Astrophys. J., 416, 399
\bibitem{daSilva99} A. da Silva  \emph{et. al.}, 2000, 
         Mon. Not. R. Astron. Soc., 317, 37
\bibitem{Springel00} V. Springel, M. White and L. Hernquist, 
        2001, Astrophys. J., 549, 681
\bibitem{Sunyaev:Zel80} R. A. Sunyaev and Ya. B. Zel'dovich, 1980, 
        Annu. Rev. Astron. Astrophys., 18, 537
\bibitem{Valageas00} P. Valageas, A. Balbi and J. Silk, 2001, 
        Astron. Astrophys., 367, 1
\bibitem{Aghanim95} N. Aghanim \emph{et. al.}, 1996, Astron. Astrophys., 311, 11 
\bibitem{Benson00} A. J. Benson \emph{et. al.}, 2001, 
        Mon. Not. R. Astron. Soc., 320, 153
\bibitem{Gnedin00} N. Y. Gnedin,  2000, Astrophys. J., 535, 530
\bibitem{Gruzinov:Hu98} A. Gruzinov and W. Hu, 1998, Astrophys. J., 508, 435
\bibitem{Knox98} L. Knox, R. Scoccimarro and S. Dodelson, 1998, 
        Phys. Rev. Lett., 81, 2004
\bibitem{Hu:White96} W. Hu and M. White, 1996, Astron. Astrophys., 315, 33
\bibitem{Jaffe:Kam98} A. Jaffe and M. Kamionkowski, 1998, Phys. Rev. D, 58, 43001
\bibitem{Gnedin:Jaffe00} N. Y. Gnedin and A. H. Jaffe, 2001, Astrophys. J., 551, 3
\bibitem{Goldberg:Spergel99} D. M. Goldberg and D. N. Spergel, 1999, 
        Phys. Rev. D, 59, 103002
\bibitem{Cooray:Hu00} A. Cooray and W. Hu, 2000, Astrophys. J., 534, 533
\bibitem{Cooray01a} A. Cooray, 2001, Phys. Rev. D, 64, 043516
\bibitem{Cooray01b} A. Cooray, 2001, Phys. Rev. D, 64, 063514
\bibitem{Komatsu02} E. Komatsu, 2002, Ph.D. thesis at 
        Tohoku University, astro-ph/0206039
\bibitem{Robertson40} H. P. Robertson, 1940, Proc. Cambridge Philos. Soc., 30, 209
\bibitem{Scannapieco00} E. Scannapieco, 2000, Astrophys. J., 540, 20
\bibitem{Spergel:Goldberg99} D. N. Spergel and D. M. Goldberg, 1999, 
        Phys. Rev. D, 59, 103001
\bibitem{Tegmark96} M. Tegmark, A. Taylor and A. Heavens, 1997, 
        Astrophys. J., 480, 22
\bibitem{Kenney:Keeping51} J. F. Kenney and E. S. Keeping, 1951, 
            \emph{Mathematics of Statistics}, Part II, 2nd ed., Van Nostrand, New York
\bibitem{Kendall:Stuart69} M. G. kendall and A. Stuart, 1969, 
            \emph{The Advanced Theory of Statistics}, Volume II, Griffin, London
\bibitem{Jungman96} G. Jungman, M. Kamionkowski, A. Kosowsky and D. Spergel, 1996, 
            Phys. Rev. D, 54, 1332
\bibitem{Knox95} L. Knox, 1995, Phys. Rev. D, 52, 4307
\bibitem{Hu99} W. Hu, 2000, Astrophys. J., 529, 12
\bibitem{Hamilton91} A. J. S Hamilton, P. Kumar, E. Lu and A. Matthews, 1991, 
         Astrophys. J. Lett., 374, 1
\bibitem{Pea:Dod94} J. A. Peacock and S. J. Dodds, 1994, 
        Mon. Not. R. Astron. Soc., 267, 1020
\bibitem{Pea:Dod96} J. A. Peacock and S. J. Dodds, 1996, 
        Mon. Not. R. Astron. Soc., 280, L19
\bibitem{Zaldarriaga99} R. Scoccimaro, M. Zaldarriaga and L. Hui, 1999, 
        Astrophys. J., 527, 1
\bibitem{Gnedin:Hui98} N. Gnedin and L. Hui, 1998, Mon. Not. R. Astron. Soc., 296,44
\bibitem{Aghanim02} N. Aghanim, P. G. Castro, A. Melchiorri and J. Silk, 2002, 
        Astron. Astrophys., 393, 381 
\bibitem{Gangui:Martin00} A. Gangui and J. Martin, 2000, 
        Mon. Not. R. Astron. Soc., 313, 323; Phys. Rev. D, 62, 103004
\bibitem{Seljak:Zaldarriaga96} U. Seljak and M. Zaldarriaga, 1996, 
        Astrophys. J., 469, 437 
\bibitem{Tegmark99} M. Tegmark, D. Eisenstein, W. Hu and A. de Oliveira-Costa, 2000, 
        Astrophys. J., 530, 133
\bibitem{Zaldarriaga00} M. Zaldarriaga, 2000, Phys. Rev. D, 62, 063510
\bibitem{Cooray:Kesden02} A. Cooray and M. Kesden, 2003, New Astron., 8, 255
\bibitem{Limber54} D. Limber, 1954, Astrophys. J., 119, 655
\bibitem{Kaiser92} N. Kaiser, 1992,  Astrophys. J., 388, 272 
\bibitem{Buchalter00} A. Buchalter, M. Kamionkowsky and A. Jaffe, 
        2000, Astrophys. J., 530, 36
\bibitem{Pollo01} A. Pollo, 2001, \emph{Where's the Matter}, 
        Proceedings of the 3rd Marseille Cosmology Conference, 
        Laboratoire d'Astrophysique de Marseille,
        eds. L. Tresse and M. Treyer, p.287  
\end{thebibliography}
\end{document}